\documentstyle[12pt, psfig, epsfig]{article}
\begin{document}

\title{{\Large \bf The formation of disk galaxies in computer simulations}}

\date{    }
\maketitle
\begin{center}
\vskip -1.5truecm

{Lucio Mayer$^{1,2}$, Fabio Governato$^{3}$ \& Tobias Kaufmann$^{4}$}\\
\vskip 0.3truecm

{\small 1-Institute for Theoretical Physics, University of Zurich, Winterthurestrasse 190, 8057 Zurich, Switzerland\\
2-ETH,  Physics Department, Wolfgang Pauli Strasse 16, CH-8093 Zurich, Switzerland, lucio@phys.ethz.ch\\
3-Department of Astronomy, University of Washington, Stevens Way, Seattle, WA 98193, USA\\
4-Center for Cosmology, Department of Physics and Astronomy, University of California, Irvine, CA 92697, USA}
\end{center}

\begin{abstract}

The formation of disk galaxies is one of the most outstanding problems in modern astrophysics
and cosmology. We review the progress made by numerical simulations carried out on large
parallel supercomputers. These simulations model the formation of disk galaxies within the current structure
formation paradigm in which the Universe is dominated by a cold dark matter component and
a cosmological constant. We discuss how computer simulations have been an essential tool
in advancing the field further over the last decade or so. Recent progress stems from
a combination of increased resolution and improved treatment of the astrophysical 
processes modeled in the simulations, such as the phenomenological description of the 
interstellar medium and of the process of star formation.
We argue that high mass and spatial resolution is a necessary condition in order to obtain
large disks comparable with observed spiral galaxies avoiding spurious dissipation of angular
momentum. A realistic model of the star formation history. gas-to-stars ratio and
the morphology of the stellar and gaseous component is instead controlled by the phenomenological
description of the non-gravitational energy budget in the galaxy. This includes  the energy injection by 
supernovae explosions as well as by accreting supermassive black holes at scales below the resolution.
We continue by showing that  simulations of gas collapse within cold dark matter halos including a 
phenomenological description of supernovae blast-waves
allow to obtain stellar disks with nearly exponential surface density profiles as those observed in 
real disk galaxies, counteracting the tendency of gas collapsing in such
halos to form cuspy baryonic profiles. However, the ab-initio formation of a realistic rotationally
supported disk galaxy with a pure exponential disk in a fully cosmological
simulation is still an open problem. We argue that the suppression of bulge formation is
related to the physics of galaxy formation during the merger of the most massive protogalactic lumps
at high redshift, where the reionization of the Universe likely plays a key role. A sufficiently
high resolution during this early phase of galaxy formation is also crucial to avoid artificial angular momentum 
loss and spurious bulge formation. Finally, we discuss the role of mergers in disk formation, adiabatic
halo contraction during the assembly of the disk, cold flows, thermal instability and other  aspects of galaxy formation, 
focusing on their  relevance to the puzzling origin of bulgeless galaxies.

\end{abstract}

~~~~~~~~~~~~~~~~keywords:astrophysics, cosmology, computer science, fluid dynamics

\section{Galaxy formation in a hierarchical Universe}

Galaxies occupy a special place in our quest for understanding the Universe. They are large
islands in a nearly empty space and contain most of the ordinary baryonic matter,  stars
and interstellar gas, that emits radiation and can thus be detected by astronomers$^1$.
Galaxies come essentially in two broad categories$^2$, those in which the luminous mass 
is arranged in a rotating disk of stars and gas, called disk galaxies or spiral galaxies because
of the presence of spiral arms of gas and stars (Figure 1), and those
in which the luminous mass is distributed in a smooth, featureless spheroidal structure with little or no 
rotation,  also known as elliptical galaxies (Figure 1).
Disk galaxies have a radial light distribution $I(r)$ that is well fit by a decaying exponential law$^3$, $I(r) \sim exp(-r/r_d)$, 
where $r_d$ is a characteristic scale length ($r_d \sim 2-4$ kpc for typical spiral galaxies). 
Indeed many disk galaxies contain also a spheroidal stellar component at their center, the stellar bulge, which has 
structural properties similar to an elliptical galaxies albeit being much smaller in size$^2$.
Both types of galaxies are known to contain dark matter, namely matter that is not traced by radiation. 
In disk galaxies dark matter clearly dominates over luminous matter by mass, as inferred from their high rotation speeds
which requires the gravitational pull of a massive and extended halo of dark matter$^4$. 
We live in a galaxy of the first kind, the Milky Way. Indeed
disk galaxies are ubiquitous in the local Universe, and also at the largest distances and
earliest epochs at which the best ground and space-based telescopes have been able to study the
 morphology of galaxies reliably$^5$. Only the most massive galaxies in 
the Universe do not posess a disk component, while this becomes progressively more dominant compared
to the spheroidal component as the mass of the galaxy decreases (Figure 2).

\begin{figure}
\centering{
\resizebox{12cm}{!}{\includegraphics{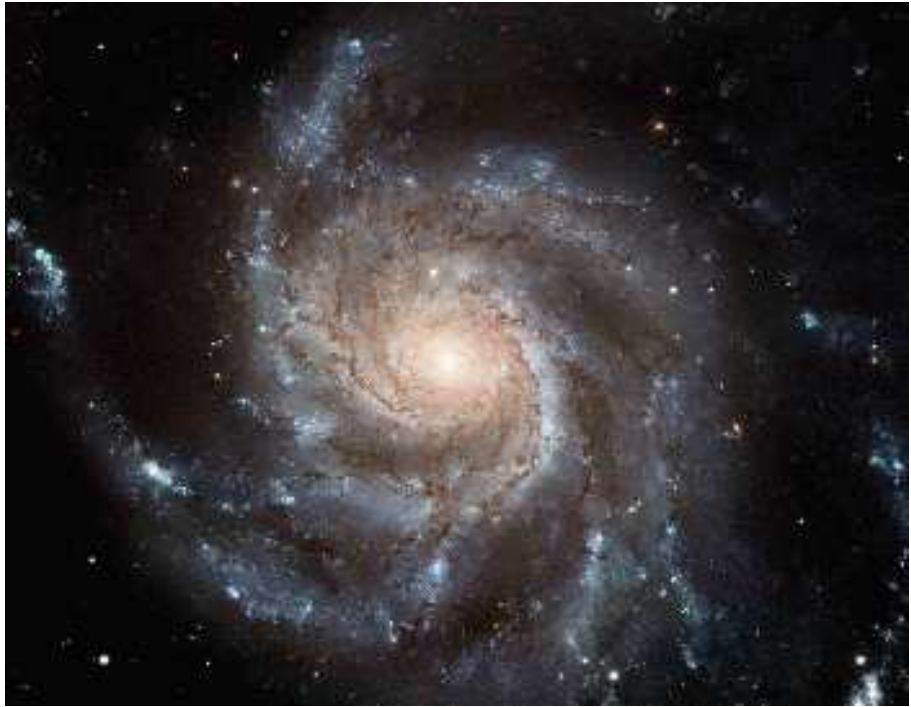}}\\
\resizebox{7cm}{!}{\includegraphics{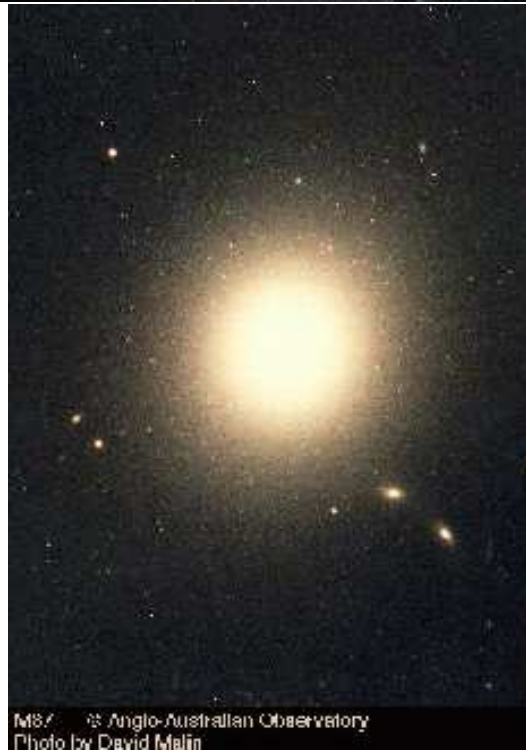}}\\
\caption{Top: a typical disk-dominated galaxy, the nearby spiral galaxy M102 in the Ursa Major constellation, 27 million light years from the Sun (the image was obtained by Chris \& Dawn Schur from Payson, Arizona at 
5150 feet elevation with an amateur telescope).
A small spheroidal bulge is visible at the center of the disk.
Bottom: a typical spheroidal
galaxy with no disk component, the elliptcal galaxy M87, located at 60 million light years from us (credits
in the picture).}}
\label{fig:feedback}
\end{figure}

\subsection{The theoretical framework}
    
The formation of disk galaxies is one of the major unsolved problems of modern astrophysics. The basic
theoretical framework states that disk galaxies arise from the gravitational collapse of a rotating protogalactic 
cloud of gas within the gravitational potential well of the dark halo$^6$. 
The gas cools via radiative processes during the collapse and eventually settles
in centrifugal equilibrium at the center of the halo potential well forming a rotationally supported gas disk   
provided that some angular momentum is retained during the collapse $^7$. These ideas were 
developed two decades ago and they still constitute the backbone
of disk galaxy formation models $^{8,9,10,11,12,13}$. What has changed dramatically 
since then is the cosmological context in which such idea is applied, which reflects
the remarkable progress that cosmology has undergone in the meantime.
After two decades of active debate there is now one cosmological
paradigm according to which the energy density of the Universe is dominated by cold dark matter and a cosmological constant,
while ordinary baryonic matter contributes only to a few percent level (an even
smaller contribution is yielded by neutrinos)$^{14}$.
This model is supported by observations of the large scale mass distribution in the Universe traced by 
galaxies  themselves$^{15}$ and by the power spectrum of density fluctuations inferred from the cosmic
microwave background radiation$^{14}$.
Cold dark matter interacts only via gravity with itself and with ordinary
baryonic matter, and is not subject to any dissipative force. 
The governing evolutionary equation for dark matter  is the collisionless Boltzmann equation that describes 
a zero-pressure fluid, also termed a collisionless fluid$^2$.
In this model, called $\Lambda$CDM (CDM stands for "cold dark matter", while $\Lambda$ is the
cosmological constant required to explain the observed acceleration of the Universe) 
structure forms hierarchically in a bottom-up fashion, starting from the amplification via gravitational instability of
primordial small density fluctuations in the dark matter$^{16,17}$.
Because of the scale-free nature
of gravity and the dissipationless nature of cold dark matter, in such a model
one expects the formation of self-similar, ellipsoidal collapsed objects, dark matter halos, at all
scales$^{18}$. The largest halos are the last to form$^{17}$.
Also, direct three-dimensional simulations of structure formation in a CDM Universe 
predict that halos of any mass and size should contain a swarm of smaller halos,
the so-called substructure$^{19,20}$, as shown in Figure 3.

\begin{figure}
\centering{
\resizebox{14cm}{!}{\includegraphics{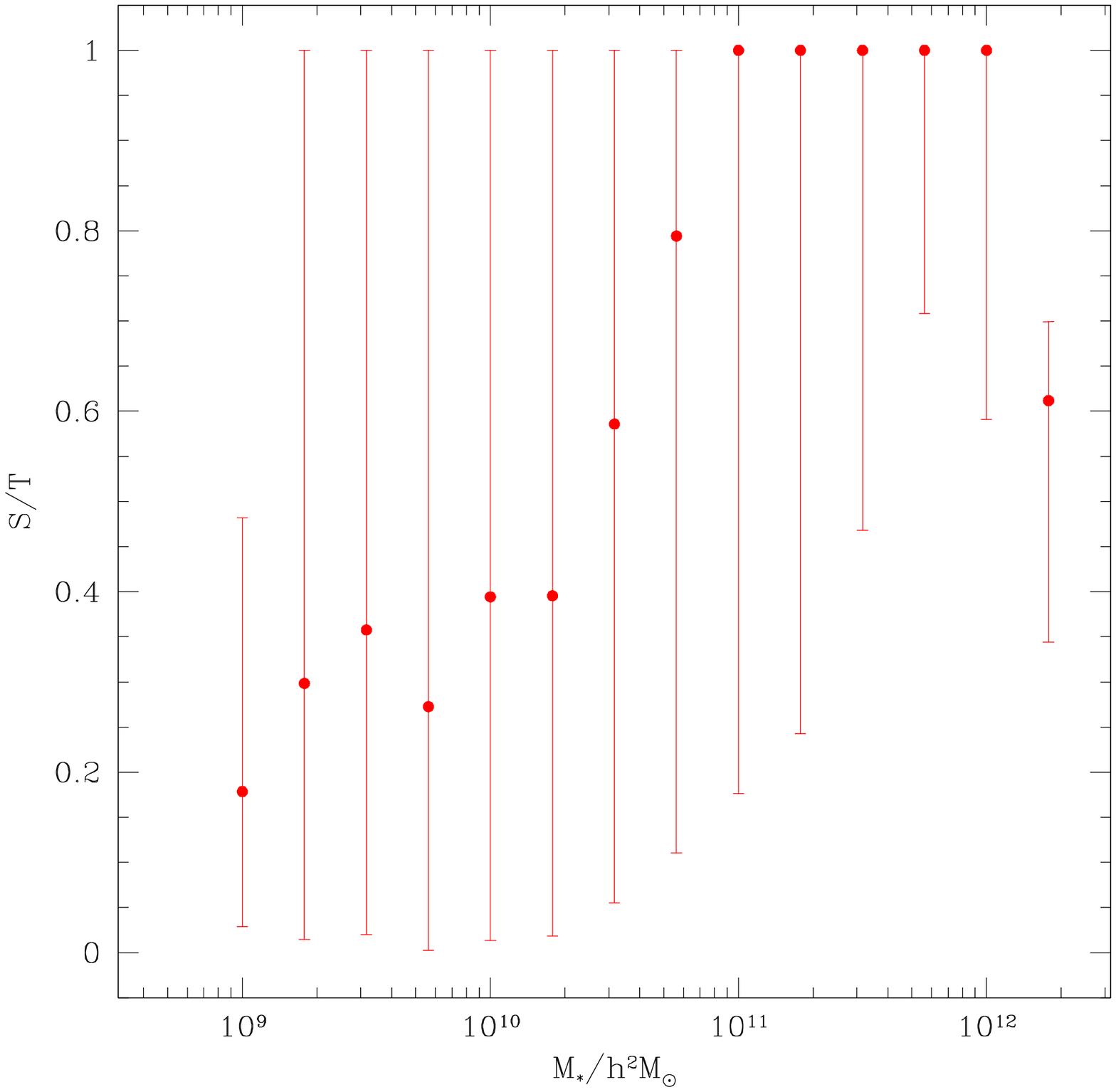}}\\
\caption{The ratio between the mass of the stellar spheroid ($S$) and the sum of the mass of the 
stellar  spheroid and the  stellar disk ($T$) as a function of galaxy mass from a galaxy sample of the
Sloan Digital Sky Surveys (SDSS)${^12}$ (2007 Blackwell Publishing Ltd). 
Red points with error bars show the median $S/T$ as a function of stellar mass together with the 10 
and 90 percentiles of the distribution.}}
\label{fig:feedback}
\end{figure}

The model predicts quantitatively the size and mass of the dark halo of a galaxy with a given measured 
rotational velocity (the rotational velocity of stars and gas probes the depth of the galaxy gravitational potential well).
For a galaxy like our own Milky Way, for example, its observed rotational velocity of $\sim 220$ km/s implies
a halo mass of about $10^{12}$ solar masses and a halo radius of about $300$ kpc$^{21}$ (by comparison, the disk 
of our galaxy contains about $6 \times 10^{10}$ solar masses of stars and about $10^{10}$ solar masses of gas$^2$).
Numerical simulations of the growth of dark matter halos also predict that the radial density profiles of
such halos diverge near the center and are well described by a power-law, $\rho \sim r^{-\gamma}$ ($\rho$ being
the density and $r$ the spherically averaged radius of the halo), with $\gamma=1-1.5$
$^{22, 23, 24, 25}$. Since dark matter dominates by mass over
ordinary baryonic matter, gas collapses within such halos, eventually forming a galaxy, because it is pulled inward by their gravitational 
attraction rather than collapsing due to its own gravity. 
In this scenario there is no such a thing as a galaxy forming in isolation,
rather structure, both dark and baryonic, builds up via continous accretion and merging of smaller systems
containing a mixture of dark and baryonic matter$^{26}$.
This highly dynamical picture emerging from the current cosmology is the main difference compared
to earlier attempts to study galaxy formation.
Halos gain their angular momentum by tidal torques due to asymmetries in the distribution of matter,
and also by acquiring the angular momentum originally stored in their relative orbit as they come together and merge into
a larger system$^{27,28,29}$.
Prevailing models have then assumed that baryons and dark matter start with the 
same specific angular momentum before the collapse begins since they are subject to the same tidal torques
$^{8}$.

\begin{figure}
\centering{
\resizebox{16cm}{!}{\includegraphics{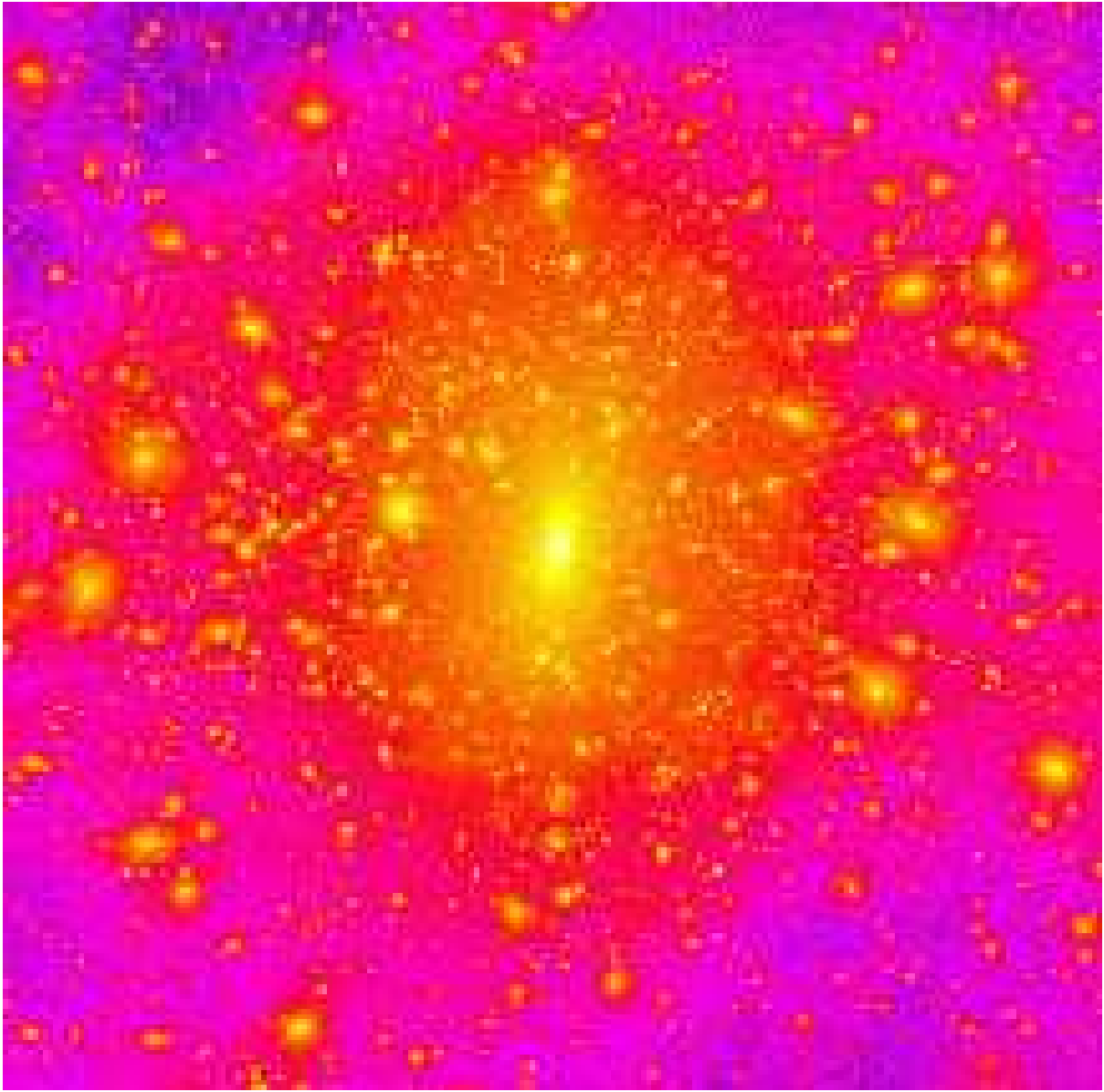}}\\
\caption{Example of a massive dark halo ($\sim 10^{14}$ solar masses) assembled by hierarchical merging in a numerical simulation
of the $\Lambda$CDM structure formation model (simulation performed by L.Mayer and F.Governato
with the GASOLINE code  $^{44}$).
Many small halos ("substructure") are orbiting inside
the larger halo and will eventually merge with it as a result of dynamical friction eroding their
orbital energy and orbital angular momentum. These dark matter lumps are the sites in which gas collapses
and forms galaxies.}}
\label{fig:feedback}
\end{figure}

\subsection{Computer simulations:the angular momentum problem}

The modern tool used to study the hierarchical
growth of structure driven by gravity and the concurrent collapse of baryons within dark halos
is represented by three-dimensional computer simulations that solve the
gravitational and hydrodynamical forces between parcels of gas and dark matter. We will discuss
the methodology employed by such simulations in the next section. For now it suffices to say that
in the most popular simulation method both the gas (i.e.
the baryonic component) and the dark matter are represented by particles so that structures
are discretized in mass and space. The evolutionary equations, such as the collisionless Boltzmann
equation for cold dark matter, the Euler equation for baryonic matter (baryonic matter is treated as
an ideal gas) and the Poisson equation, which holds for both, are solved for such discrete representation
of physical reality.
Available methods to discretize physical variables and governing equations are constructed in such a way 
that they should converge to the exact continuum solution for an infinite number of particles. 
As we will see in the next section, discretization itself, along with other aspects of the current methods, can 
introduce spurious effects in the computer models.
Simulations take advantage of large parallel supercomputers in which hundreds of processing units are
used simultaneously to compute the forces and advance the system to the next timestep.

One important prediction of simulations of a CDM Universe 
is that halos have a rather universal value of the angular momentum (per unit mass) 
at any given epoch quite irrespective of their precise mass assembly history. This is parameterized 
via the dimensionless spin parameter $\lambda = J E^{1/2}/GM^{5/2}$, where $E$, $M$ and $J$ are
the total energy, mass and angular momentum of the dark halo ($G$ is the gravitational constant).
One can show that $\lambda$ is proportional  to the ratio between the rotational kinetic energy and 
the kinetic energy in disordered motions associated with the halo. 
Halos have a universal distribution of spin parameters, which peaks at a value $\lambda \sim 0.035$
$^{29}$.

Simple spherical one-dimensional models that study disk formation in an isolated 
CDM halo (namely a halo that does not interact or merge with other halos)
predict that  the size of disks resulting from the infall and collapse of baryons matches the size of observed disks in galaxies
very well$^8$. 
The models use mainly two inputs, both coming
from cosmological simulations, the halo density profile, which is related to the gravitational pull
that drives the gas collapse, and the initial specific angular momentum of gas as 
implied by the typical values of the spin parameter. They further assume that angular momentum is conserved during the
 collapse. Hence this result is simple and remarkable at the same time; it says that CDM halos have the right amount of 
angular momentum to form observed disk galaxies.

For more than a decade researchers have tried to reproduce the latter result with fully three-dimensional
computer simulations but have run into several problems. It was soon realized that, once the hypothesis
of isolation is removed and hierarchical merging is accounted for, angular momentum can be lost
by the gas to the dark matter due to a process known as dynamical friction$^{30,2,31}$.
During mergers, 
previously collapsed clumps of gas and dark matter fall into a larger dark halo and suffer a drag force as they move
through the latter. The loss of angular momentum caused by the drag force, called "dynamical friction",
is more effective when the gas is distributed into cold and dense lumps rather than
being smooth and extended$^{32}$.
But gas is expected to be clumpy  in a model with collisionless cold 
dark matter in which collapse can occur at all scales, and large halos grow by accreting smaller halos 
which bring their own dense collapsed gas. As a result, early simulations$^{32}$ 
were obtaining improbable small disks with ten times less angular momentum than real ones.
Two types of solutions for this "angular momentum problem" have been considered since then.
The first is very drastic and calls for
revising the cosmological model itself. Alternative models in which the dark matter has a non-negligible thermal
velocity rather then being "cold" would produce collapsed systems only above a characteristic scale
because the thermal jittering will tend to smear out short-wavelength density perturbations$^{33}$. These warm dark
matter models (WDM) behave like CDM on large scale, thus maintaining its succesful features.
The reduced clumpiness of dark matter halos in the WDM model implies that baryons are smoothly
distributed rather than arranged in previously collapsed dense lumps when they fall into
large galaxy-sized halos, and therefore lose less angular momentum by dynamical friction$^{33,34}$.
The second, less exotic possibility, is that baryons do not just follow the merging hierarchy imposed by
dark matter but somehow decouple from it  and remain much smoother. This could happen if the
thermal energy content of baryons was enough to resist gravitational collapse, at least up to
some critical mass scale. This way a fraction of the gas that would have entered a halo in dense
clumps within smaller halos would instead enter with a smooth distribution, perhaps avoiding catastrophic loss of angular momentum.
Various plausible astrophysical mechanisms can be responsible for increasing the thermal energy
content of the baryons, for example the energy injection by supernovae explosions and the ambient
radiation field produced by stars, accreting black holes or also external galaxies.
There is, however, a third possibility. This
is that the baryons clump excessively in computer simulations because the numerical methods adopted  
can introduce artificial loss of angular momentum. As we argue in this report, a 
solution lies probably in a combination of the latter two proposals, with no need of revising
the standard cosmological structure formation model.

\begin{figure}
\centering{
\resizebox{18cm}{!}{\includegraphics{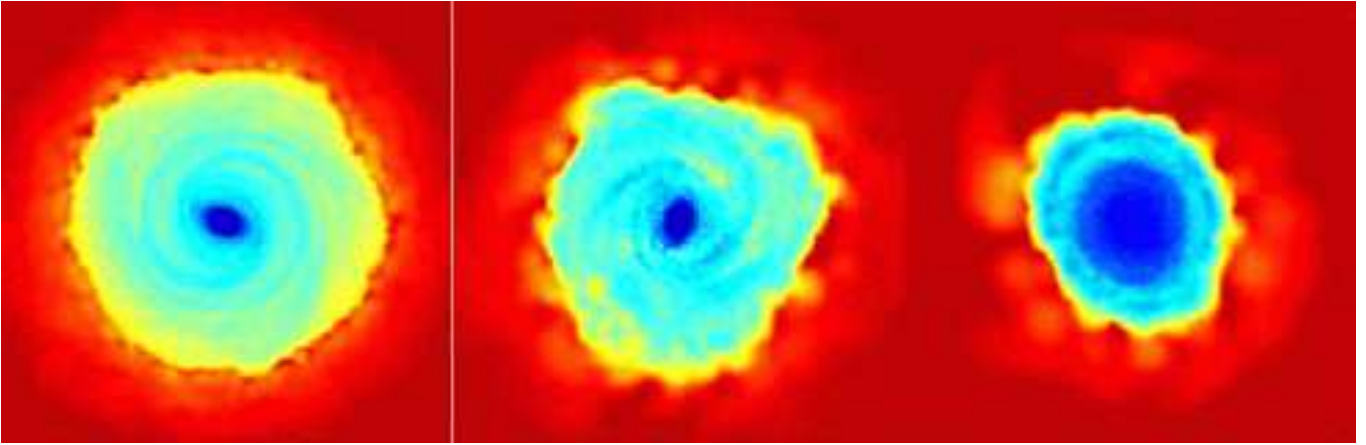}}\\
\caption{ Disk size as a function of mass resolution in numerical simulations of disk formation in an isolated
CDM halo$^{48}$ (2007 Blackwell Publishing Ltd).
The three panels show density maps of gas in a slice through 
the centre of the gaseous galactic disk after 5 Gyr;  the gas mass resolution decreases from the left to the right
by about a factor 8 each snapshot (the maximum resolution is 1 million gas particles). The box side length 
is 20 kpc for every panel. At all resolutions disks show asymmetries such as central bar-like structures
and spiral arms. The bar, however, is a strong and long-lived feature only for sufficiently high spatial resolution
(set by the gravitational softening)$^{48}$.}}
\label{fig:feedback}
\end{figure}

The next two sections will be devoted, respectively, to the role of numerical effects in disk formation
simulations, and to the modeling of gas thermodynamics and star formation in the simulations. We will
then show how the structure of simulated disks is affected by different models of thermodynamics
and star formation. Finally, we will summarize the current status of the field and  the major problems that 
remain to be solved, including the puzzling origin of disk galaxies without a bulge.
We will attempt to recall the most important contributions by the various 
groups actively involved in this field of research while at the same time covering in
more detail some recent results of the research group to which the authors of this report
belong.

\section{Modeling issues and numerical effects in computer simulations of disk formation}

\subsection{Numerical methods}

Before discussing numerical effects in cosmological disk galaxy formation simulations we should
keep in mind that until the time of writing this field of research
has been dominated by one numerical method, smoothed particle hydrodynamics (SPH)$^{35,36,37,38}$, 
in which particles are tracers of average fluid quantities such as density, velocity and 
temperature associated with  a finite volume of the fluid. The fluid
is thus discretized by means of particles and the calculation of the fluid-dynamical equations is carried
out in a Lagrangian fashion. The Euler equation for an inviscid ideal gas is solved rather than the more
general Navier-Stokes equation. An artificial viscosity term is introduced in the hydrodynamical force equation
and in the internal energy equation to compensate for some artifacts resulting from the discrete representation of 
the fluid equations, namely to avoid particle interpenetration and damp spurious oscillations in shocks$^{39}$.
Although it is introduced for these good reasons, artificial viscosity also causes some unwanted
numerical effects, such as damping the angular momentum of the fluid.
One of the reasons behind the dominance of the SPH   method in this field is that it couples
naturally with the most efficient and accurate methods to compute gravitational forces in the fluid
as well as in the collisionless dark matter component, the so called treecodes$^{37}$.
Treecodes are
an approximate but fast and accurate way of solving the N-Body problem$^{40}$.
In treecodes gravitational
forces between all particles, dark matter and baryons, are solved via a type of multipole expansion
of the gravitational field which reduces to individual particle interactions only at short distances.
Eulerian techniques
that solve gravity and the fluid equations on a fixed or adaptive grid have been less used in this field
of research because they have been generally slower and less accurate when it comes to compute gravity.
However, the scenario is changing rapidly
with the appearance of several fast, parallel adaptive mesh refinement codes (AMR) that are beginning
to have an impact in  studies of galaxy formation within a cosmological context$^{41,42,43}$.
Restricting ourselves to the current 
particle-based methods, we will
now briefly review the most important numerical artifacts that can severely affect the angular momentum
content and structure of disks forming in the simulations. Many of the results that we will discuss in
greater detail in this and in the following sections were obtained with the tree+SPH code GASOLINE$^{44}$ 
in simulations performed on large parallel supercomputers.

\subsection{Numerical effects: two-body heating}

One major problem of all particle-based
simulations, both hydrodynamical and collisionless, is numerical two-body heating. Two-body heating is the
spurious increase of kinetic or thermal energy of a particle representing gas, stars or dark matter due
to a collision with another particle. Particles behave as gravitating point masses and therefore can
undergo strong gravitational accelerations in close encounters with consequent large transfers of momentum.
Such an effect is clearly an artifact of the discrete representation of a continuum by means of particles.
For gas particles, such spurious large accelerations can be partially
compensated by pressure or artificial viscosity that tend to deflect particles as they approach
one another. In order to partially overcome this problem
for all types of particles, gravity is decreased at small distances thanks to the
introduction of gravitational softening$^{45}$.
Softening the gravity field at scales of order the
interparticle separation is consistent with the fact that individual
particles should represent a fairly large sub-volume of the hydrodynamical fluid or collisionless
continuum (e.g. the dark matter component) rather than real point-masses (in typical
simulation dark matter and baryonic particles can indeed weight $10^6-10^7$ solar masses)
However, in cosmological simulations of the CDM model
dark matter particles are typically much heavier than gas and star particles because 
dark matter accounts for most of the mass. As a result
massive dark matter particles can transfer significant kinetic energy during gravitational
encounters with other particles despite the presence of gravitational softening.

\begin{figure}
\centering{
\resizebox{8cm}{!}{\includegraphics{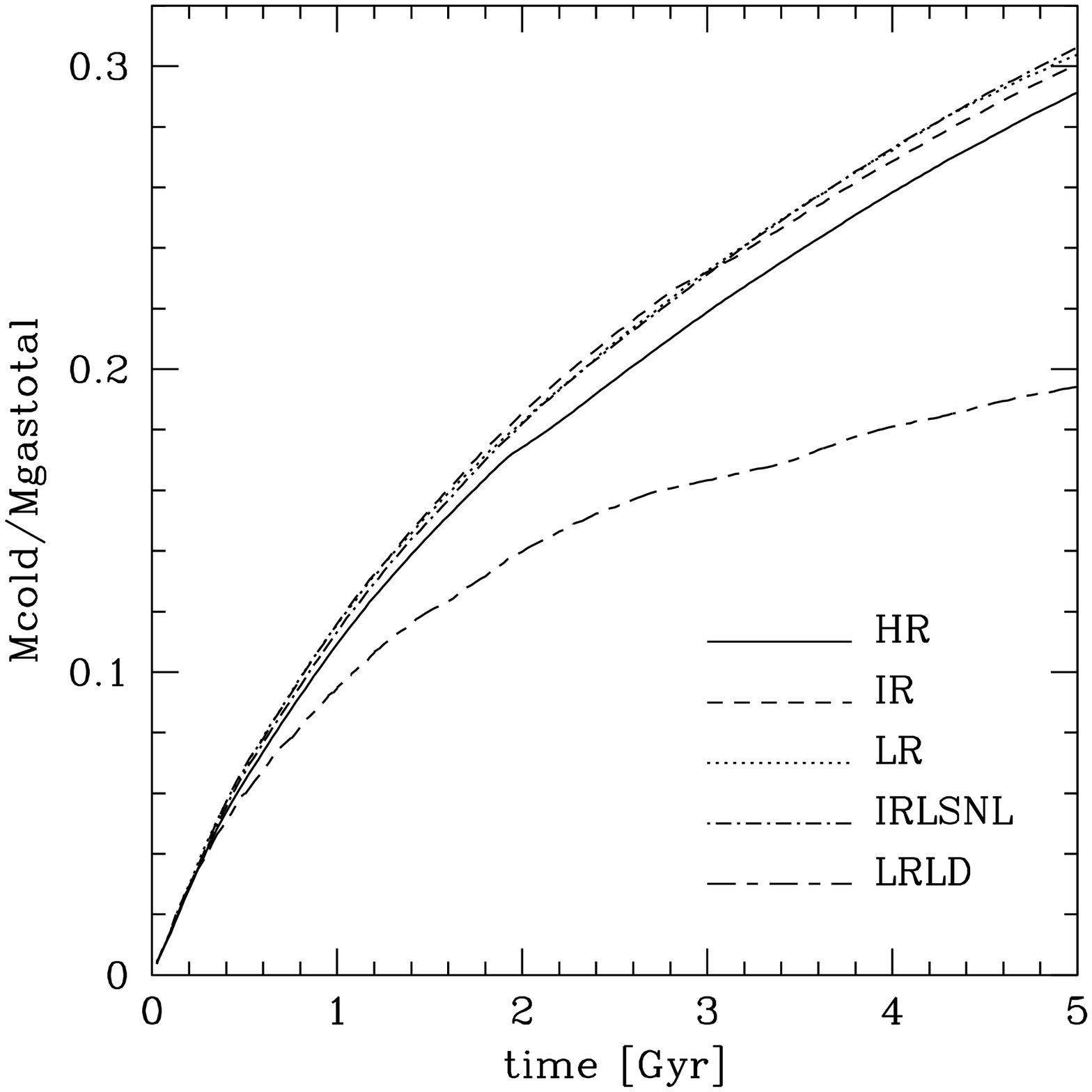}}\\
\resizebox{8cm}{!}{\includegraphics{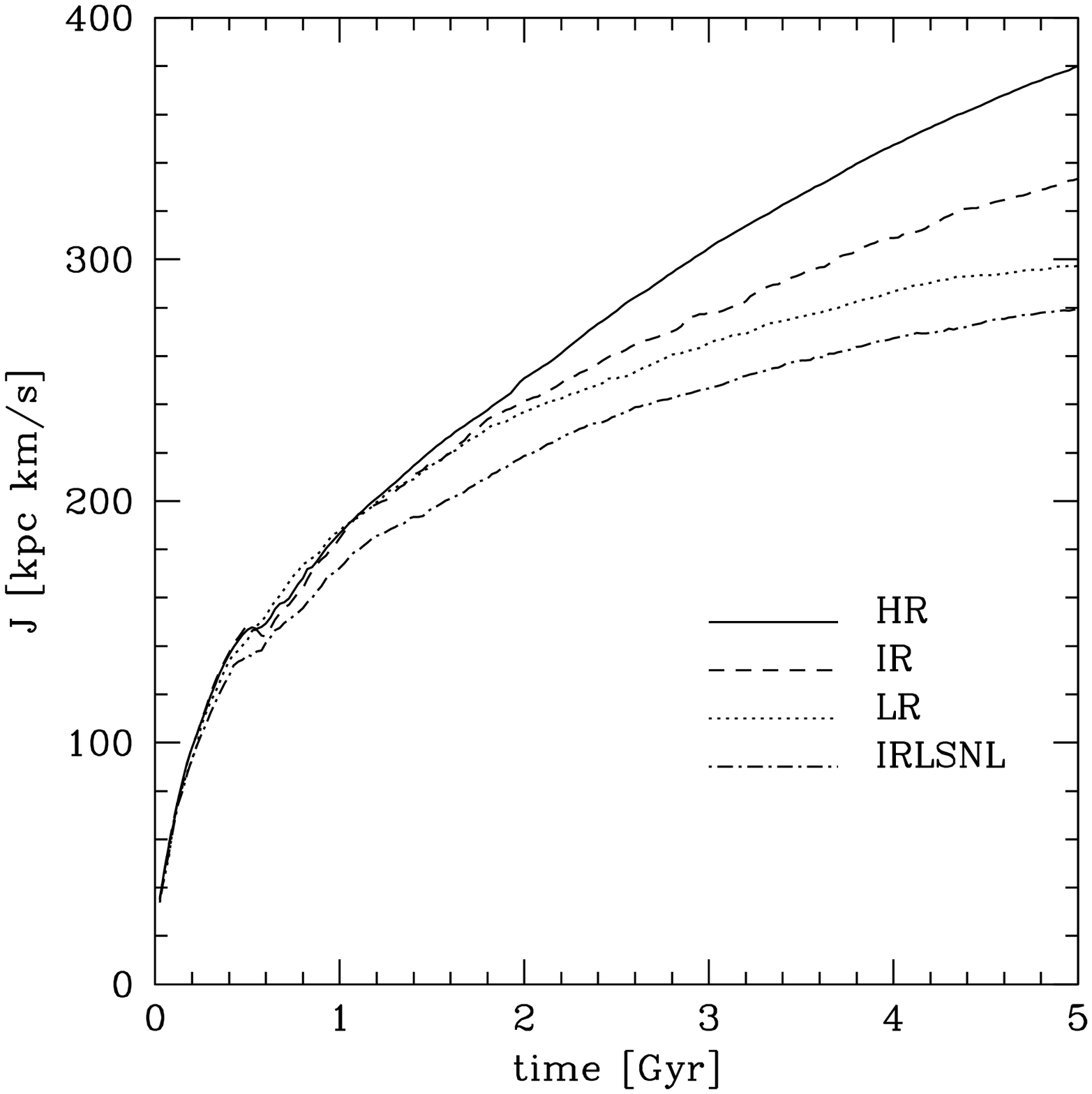}}\\
\caption{Top: growth of the  disk mass (expressed relative to the total gas mass
within the dark halo) as a function of resolution$^{48}$ (2007  Blackwell Publishing Ltd).
Bottom: Evolution of the  specific angular momentum in the disk as function of resolution${48}$. 
The resolution of the simulations is increasing from 
LRLD ($3 \times 10^4$ gas matter particles and $2.5 \times 10^3$ dark matter particles)
to LR ($3 \times 10^4$ dark matter and gas particles), IR ($9 \times 10^4$ dark matter and 
gas particles) and  HR ($5 \times 10^5$ gas matter particles and $10^6$ dark matter particles)
; IRLSNL differs from the other runs in the prescription
of artificial viscosity. See Table 1 in $^{48}$ for details on the simulations.  
}}
\label{fig:feedback}
\end{figure}

This spurious transfer of energy in two-body encounters increases
the kinetic energy in random motions because any component of the velocity can
be boosted as a result of a given encounter. Numerical experiments have shown that rotationally supported, thin stellar disks
can be gradually degraded into a thick spheroidal distribution because the random velocity of the baryonic particles
becomes increasingly more important compared to ordered rotation$^{46}$.
If two-body heating is moderate and a recognizable disk
component survives the angular momentum of the disk along the original
axis of rotation still decreases as a result of the randomization of velocity vectors$^{46}$. 
Hence two-body heating induces artificial angular momentum loss.
Gas particles suffer an increase of temperature as a result of two-body heating$^{47}$.
This affects the radiative cooling efficiency of the gas because
the cooling rate is a function of temperature$^{47}$.
The increase of temperature occurs because the kinetic energy gained in two-body collisions 
is thermalized by artificial viscous dissipation.
The only way to reduce these effects is to reduce the graininess of the mass distribution, which
is only achieved by increasing the number of particles used in the simulation. This of course calls
for more computing power. 

The reduction of spurious effects induced by two-body heating with
increasing resolution has been tested systematically by means of toy-models that represent
an isolated, already assembled galaxy$^{34,46}$.
This type of models is idealized but allows to gain insight
into the phenomenology of the much more complex cosmological simulations where many interacting objects are
simultaneously modeled. Such studies have
indicated that $> 10^5$ particles are required in the dark matter of a single galaxy to keep 
the spurious kinetic energy increase to levels $< 10 \%$.
In typical cosmological simulations many galaxies are followed at the same time and the
latter becomes a tough resolution requirement.

\begin{figure}
\centering{
\resizebox{16cm}{!}{\includegraphics{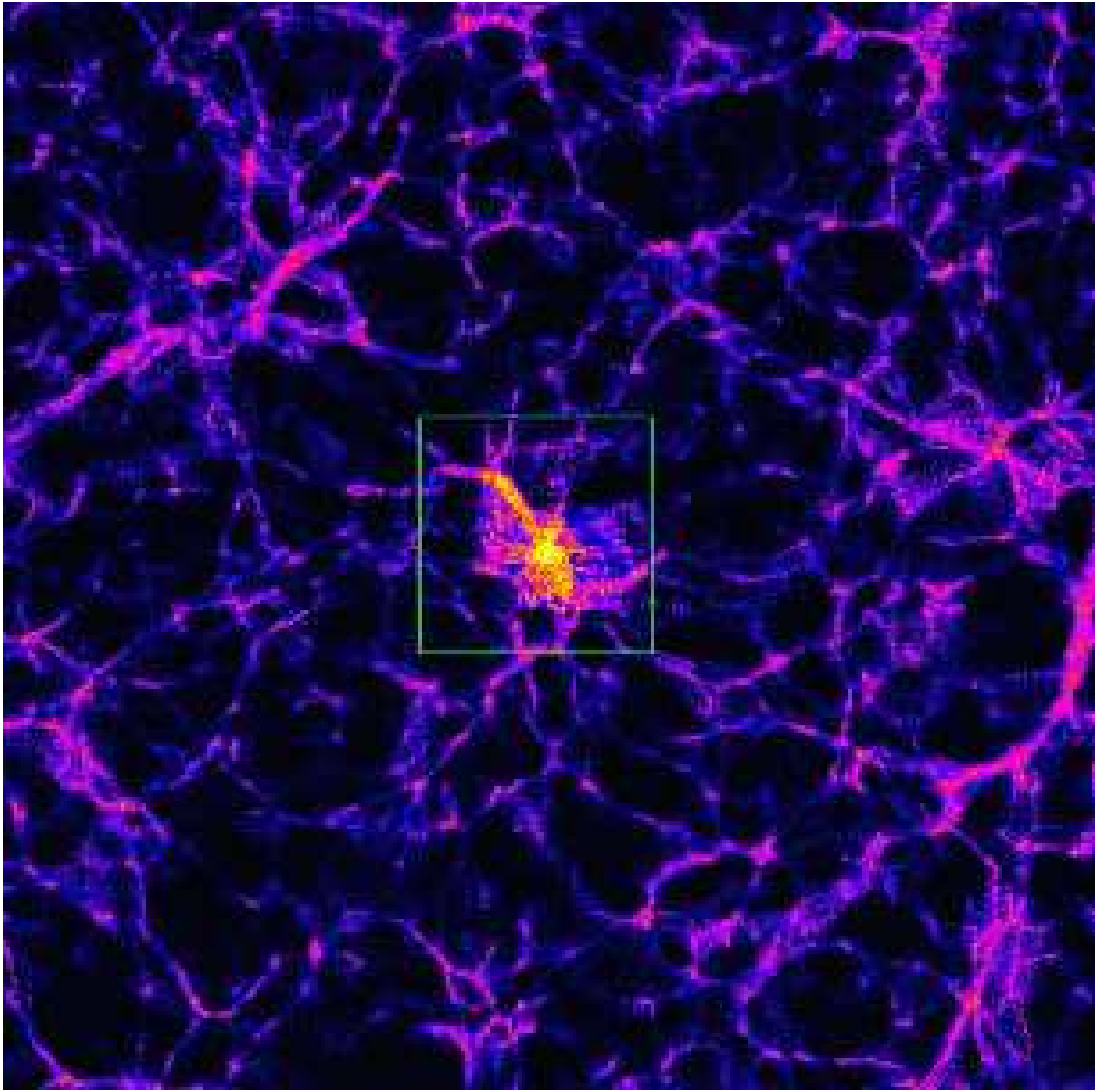}}\\
\caption{Example of renormalized cosmological volume$^{34}$. The spatial 
distribution of dark matter particles during the simulation is shown, color coded in density. The 
region marked in green has been resampled at higher resolution compared to the rest of the volume 
(whose total size is $100$ Mpc) to follow the formation of a galaxy-sized object inside it.}}
\label{fig:feedback}
\end{figure}

\subsection{Numerical loss of angular momentum in the baryonic component}

When the resolution in the dark matter component is high enough to overcome two-body heating there 
are other numerical artifacts that can affect the formation and evolution of
galaxies in the simulations, including the size and angular momentum of the disk. These are related to
how well the gas component is resolved, namely on the number of gas particles. 
and is once more best studied using idealized numerical experiments with 
isolated galaxy models. These experiments show
that gas particles can lose angular momentum as they collapse within the dark matter halo
before they settle into a centrifugally supported orbit and join the disk$^{49,50,48}$.
Spurious angular momentum loss can happen for various reasons; (1) artificial viscosity; (2) 
the interaction between particles with significant temperature
difference. The disk edge, where cold particles already belonging to the disk are
close to hotter particles still in process of cooling and collapsing from the outer part of the halo.
The cold particles suffer an artificial drag from the hot particles as a result of an erroneous
estimate of the pressure gradient performed by the standard SPH method 
$^{51,52,50}$, and shrink to a smaller radius;
(3) the disk can be quite asymmetric at low gas resolution and because of this suffers a strong 
gravitational torque from the surrounding halo, which then extracts its angular momentum.
This latter effect is very subtle because disks might become asymmetric as a result of internal,
physical evolution of the mass distribution. Some observed galaxies indeed have asymmetries, such as bars and
oval-shaped disks. Nevertheless the asymmetry seen in the low-res simulations is artificial because
it partially disappears as the resolution is increased (Figure 4 shows that the disk becomes
rounder as the resolution is increased).
The overall result of numerical experiments at varying resolution 
is that, while the amount of collapsed mass in the disk converges rapidly with resolution (Figure 5),
angular momentum loss becomes negligible only when the number of gas particles in an
individual galaxy model approaches $10^6$ (Figure 5).

\begin{figure}
\centering{
\resizebox{10cm}{!}{\includegraphics{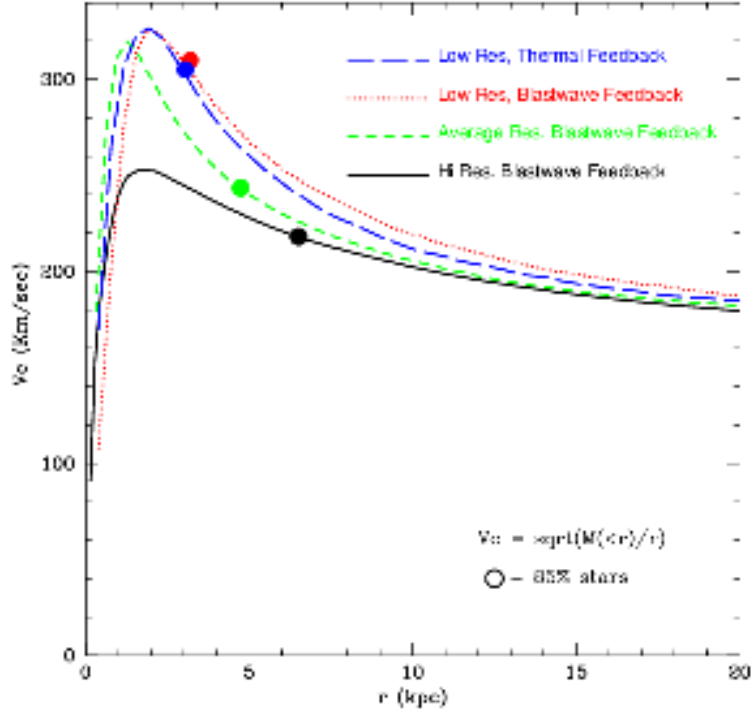}}\\
\resizebox{10cm}{!}{\includegraphics{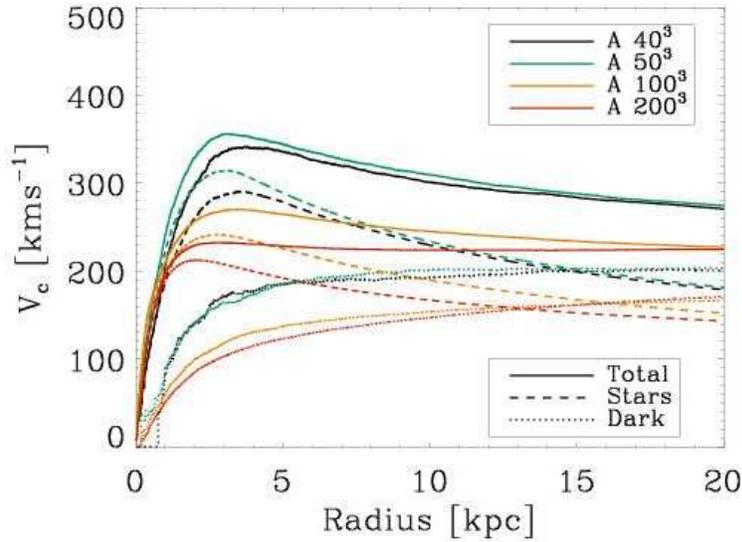}}\\
\caption{ Top: 
Rotation curve in cosmological disk galaxy as a function of resolution and feedback model$^{55}$.
See section 3.4 for a description of feedback models. The resolution of the
simulated galaxy varies from $\sim 10^4$ (low-res) to $\sim 10^5$ (average-res) to $\sim 10^6$ SPH
and dark matter particles within the virial radius. The open dots mark the radius at which $85\%$
of the stars are located.
Bottom: Rotation curves of massive galaxies formed in cosmological simulations 
for varying resolution$^{61}$ (reproduced by permission of the AAS, 2007).
The cosmological box is simulated with
four different numerical resolutions: $40^3$, $51^3$, $100^3$, and $200^3$ SPH particles and collisionless
dark matter particles. Note how the rotation curves become increasingly flat as
the resolution increases.
In both cases the rotation curve is 
calculated including all components, dark matter, stars and gas, in the computation of the mass.}}
\label{fig:feedback}
\end{figure}

Once the disk has formed, artificial viscosity can continue to degrade angular momentum, 
especially near the center of disks where velocity
gradients become very steep and the relative motion of the particles is poorly modeled.
Indeed without efficient heating by supernovae feedback or by other mechanisms, the inner 
disk always becomes very dense and loses angular momentum, even with
millions of SPH particles$^{48}$. Gravitational instability in the
gaseous or stellar component of the disk (this arises when the disk becomes quite massive 
as more gas is accreted from the halo) can also generate transport
of angular momentum via non-axisymmetric structures such as bars and spiral arms$^{2}$ (see Figure 4),
which are ubiquitous in observed galaxies$^{53}$. 
Bars can transport angular momentum outward very efficiently$^{2}$.
As a result they can flatten the disk density profile by pushing out the gas and/or the stars 
just outside the bar region$^{54}$. This affects the final disk size, with a tendency
to increase it relative to the case where no bar forms. Unfortunately, even
this kind of angular momentum transport depends on resolution; this time the 
spatial resolution of self-gravity, which is set by the gravitational softening
length,  has to be high enough to capture the wavelength of the non-axisymmetric modes of
the density field responsible for bar formation$^{48}$.

\subsection{Resolution issues in cosmological disk formation simulations}

Cosmological simulations of disk formation follow the nonlinear development of structure in 
a fairly large volume, of order $50$ Mpc$^3$, which is much larger than the volume occupied 
by an individual galaxy-sized halo. The reason of this apparent redundancy is that the large
scale tidal field needs to be included in order to properly compute the tidal torques that
generate the angular momentum of the galaxy-sized halos. The larger
volume implies more mass to sample and therefore implies that  the resolution requirements 
indicated in the previous section will be much harder to meet. To overcome this problem
a technique has been developed, and constantly improved in the last decade or so,
to increase the resolution in a selected region within a large cosmological volume. This way
the halo of a single galaxy can be resolved by up to a million dark matter and gas particles 
$^{34,55,56,57}$. This is the so-called renormalization technique (Figure 6).
In this technique, the computational volume includes regions with varying mass and spatial
resolution; the masses of the particles,  and therefore their mean separation,
increase with the distance from the selected
object, for example a single galaxy-sized halo. This way most of the particles are placed
in the region of interest, optimizing the use of computer time.
Thanks to the latter technique, more evidence has been gathered on the effect of resolution 
directly in cosmological simulations. The analysis of the renormalized simulations
confirms that disk size and angular momentum 
increases as the resolution of both the dark matter and baryonic component is increased. 
Simulations with hundreds of thousands  baryonic and dark
matter particles do produce stellar disks whose size is indeed comparable with that of real galaxies
$^{58,59,34,55,56,60}$. The most recent and highest resolution ($>10^6$ particles per galaxy) of these simulations 
even produce high angular momentum gas disks that extend well beyond the stellar
component, as observed in most spiral galaxies (Figure 8). 
However, a spheroidal bulge component always forms at the center of the galaxy (Figure 7, 8, 9) 
The bulge comprises low angular 
momentum material brought by several mergers early in the assembly history of the system$^{34,58}$.
It is too dense and massive compared to that seen in typical disk galaxies. On the other end, such a dense and massive 
bulge is not very different from that of at least {\it some} galaxies, such as  the closest 
nearby spiral galaxy, Andromeda. It is conceivable that the high density and large
mass of the bulge might be at least partially caused by artificial loss of angular
momentum during the early stages of galaxy assembly. 
In fact, even in the highest resolution renormalized simulations the progenitors 
of the final galaxy have only a few thousand particles when the bulge forms, ten billion years 
before the present epoch$^{55}$, a resolution well below that recommended
by numerical experiments with isolated galaxies$^{48}$.

\begin{figure}
\centering{
\resizebox{16cm}{!}{\includegraphics{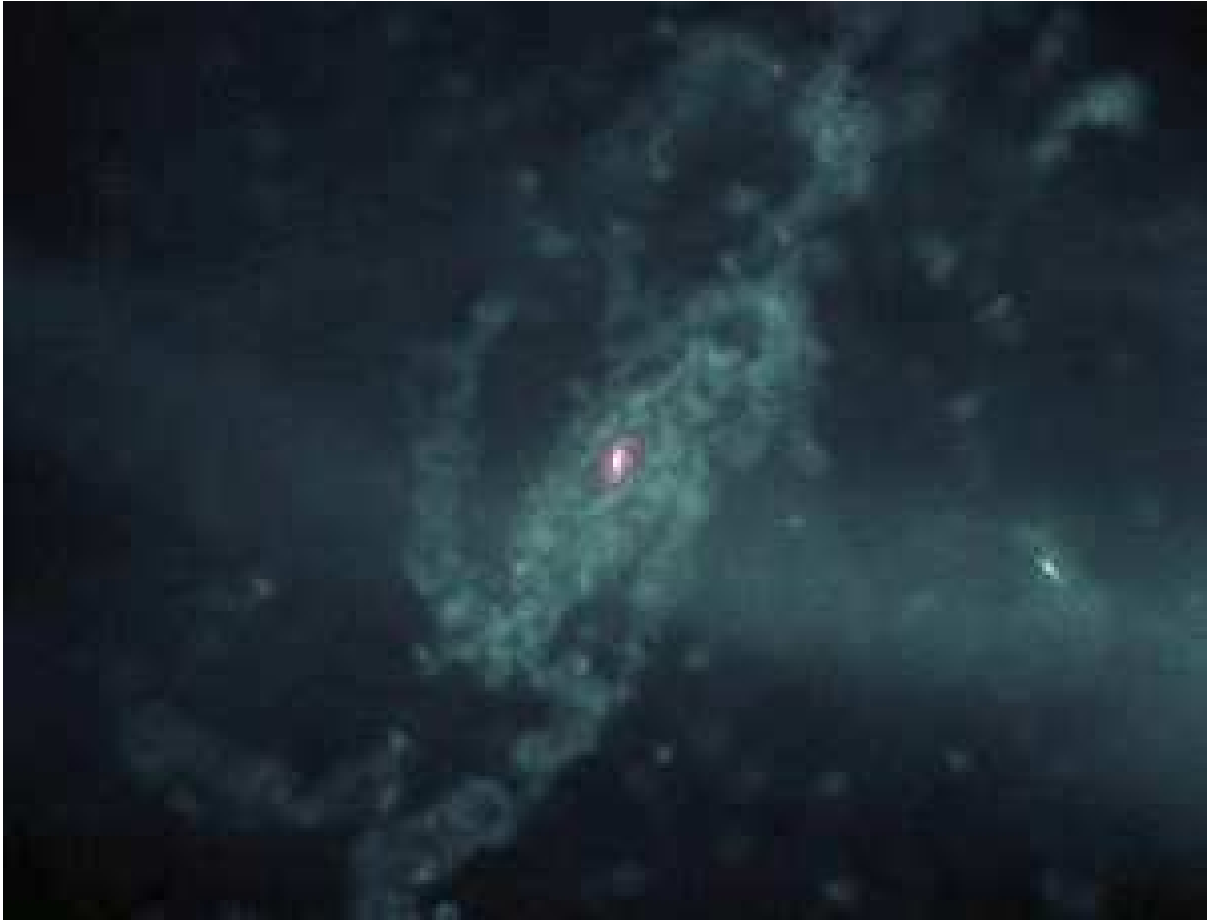}}\\
\caption{Snapshot of a high-resolution galaxy formation simulation at a time close to the present epoch 
(the resolution is comparable to the highest resolution runs published by our group$^{55}$ ). The gas is shown in green
and the stars are coloured differently based on when they formed. It can be seen that most of the stars
are concentrated in the central region while gas is a arranged in an extended disk component with radius
about 50 kpc (the whole image is about $200$ kpc on a side)}}
\label{fig:feedback}
\end{figure}

It is encouraging that the central concentration of mass,
like the disk size, also decreases as the resolution is increased. This can be quantified
by measuring the rotational velocity of the baryonic material as a function of radius, hence deriving the
"rotation curves" that observational astronomers use to probe the dynamics of real galaxies (Figure 7).
The degree of rotation is determined by the depth of the gravitational potential well; assuming spherical
symmetry, and that all the kinetic energy of the baryons is in rotational form, the rotational velocity at any given 
radius in the galaxy will be given by $V_{rot} =
\sqrt{(GM/R)}$ ($M$ is the mass, $R$ is the radius and $G$ is the gravitational constant).
$V_{rot}$ tends to 
increase towards the center if a massive central bulge component is present. Likewise,  it becomes flatter near the center 
as the mass of the central bulge diminishes. Figure 7 shows the rotation curves obtained  by two
different groups using two different numerical codes$^{55,61}$. It demonstrates  that the rotational
velocity decreases with increasing mass resolution, as expected if the spurious angular
momentum loss of the baryonic material decreases with increasing resolution. From the Figure it appears
that the rotation curve is also affected by other factors, namely the presence or absence
of heating by supernovae feedback that will be discussed in the next section. Stronger heating
will in fact counterbalance gas cooling; it will thus reduce the amount of cold gas that collapses
to the center of the progenitor halos forming the stars that will be later incorporated into the
bulge.

\section{The role of ISM physics and star formation: sub-grid models in simulations}

\subsection{The physics of the ISM}

Despite the large amount of dark matter that they contain, disk galaxies are identified via their baryonic 
component, namely gas and stars.
There is general consensus that the thermodynamics of the interstellar medium (ISM), the gaseous component in
the disks of  galaxies, is a crucial aspect of galaxy formation and evolution. Stars indeed form out of the ISM, 
being the end result of the gravitational collapse of the  densest regions of clouds made of molecular gas.
The interstellar medium in our Galaxy is multi-phase$^{62}$. 
A minimal ingredient of a realistic model of the ISM should account for the four main phases found in the disk
of a galaxy, which, in order of increasing density, are the 
the warm ionized medium (WIM, $\rho \sim 0.5$ atoms/cm$^3$, $T \sim 10^4$ K), the warm neutral medium (WNM, $\rho \sim 0.5$ atoms/cm$^3$,
 $T \sim 10^4$ K) , 
the cold neutral medium (CNM, $\rho \sim 10-50$ atoms/cm$^3$,  $T \sim 10^3$ K
which comprises clouds of atomic hydrogen) and the cold molecular
phase (mostly clouds of  molecular hydrogen, $\rho > 300$ atoms/cm$^3$,  $T \sim 10$ K ), with the first three phases being in approximate 
pressure equilibrium$^{62}$.
A hotter, diffuse phase also surrounds the disk (hot intercloud medium, $\rho \sim 3\times  10^{-3}$ atoms/cm$^3$,  $T \sim 10^5$ K
) and is likely constantly
fed by supernovae explosions that inject large quantities of thermal energy and momentum in the ISM.
The different phases are the result of thermal balance between radiative heating and cooling at different densities 
and, at the same time, thermal instability (perhaps coupled with gravitational instability) that determines the 
emergence of the two densest phases, the CNM and the molecular phase$^{62}$. Molecular gas
is formed and destroyed via a number of microscopic interactions involving ions, atoms and catalysis on dust grains.
These processes become biased towards the formation rather than towards the destruction 
of molecular hydrogen only at densities $> 10$ atoms/cm$^3$. 
A great deal of energy in the interstellar
medium is non-thermal; this turbulent energy, which is essentially observed as random gas motions of clouds and inside clouds
is supersonic, being several times larger than the thermal energy at the scale of giant molecular complexes. Turbulent kinetic
energy is thought to be the main agent that supports the largest molecular clouds ($> 10^5$ solar masses) against global 
collapse $^{63}$. The partial suppression
of gravitational collapse owing to turbulent support also explains the low efficiency of star
formation in our Galaxy (only a few percent of the molecular gas mass present
in the Milky Way appears to be involved in forming stars).
Magnetic fields also play an important role in resisting gravitational collapse at scales larger than $0.1$ pc, while below
this scale ambipolar diffusion and Ohmic dissipation give way yield to the action of gravity$^{62}$.

\begin{figure}
\vskip 12truecm
{\includegraphics{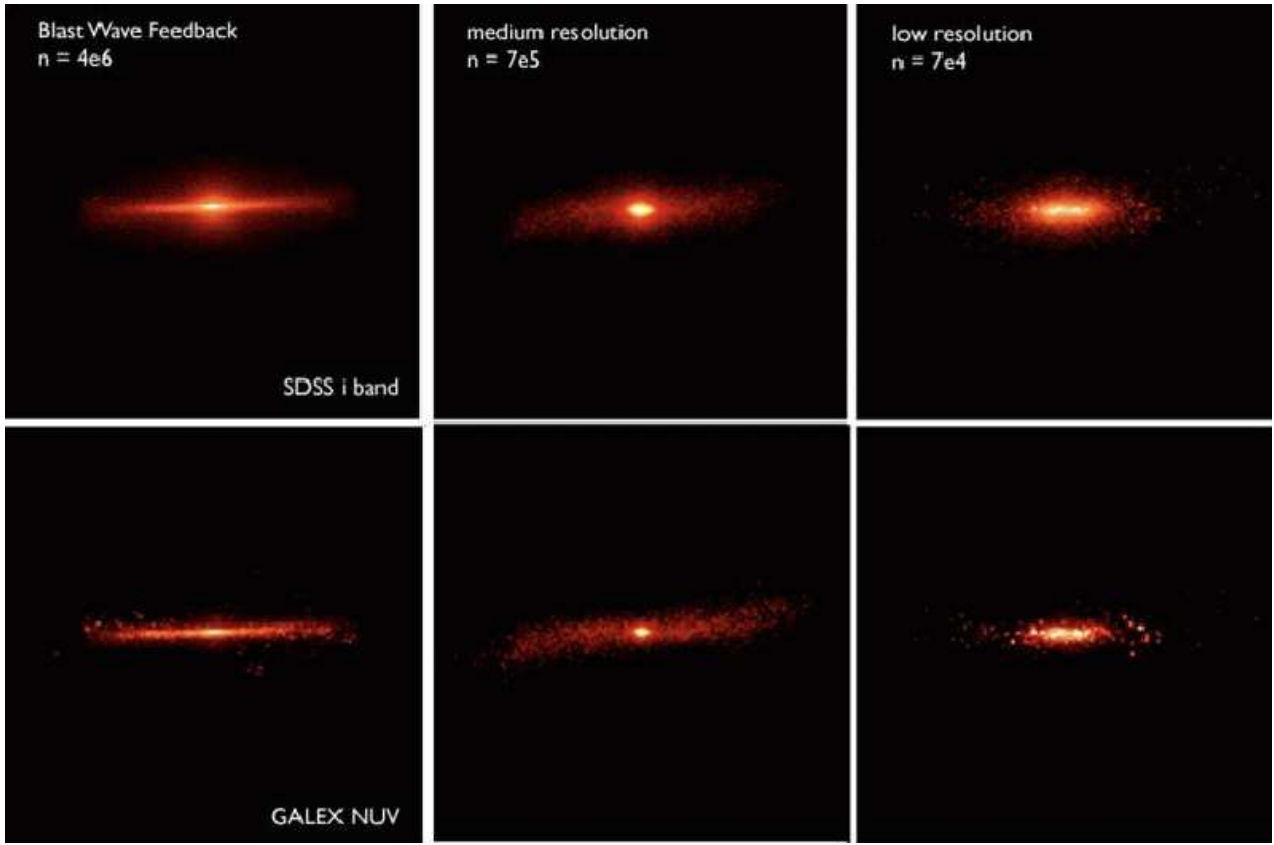}}
\caption[]{\small 
Disk size and morphology in cosmological simulations of a Milky-Way sized galaxy$^{55}$
with varying resolution, performed with the blast-wave feedback sub-grid model described in
section 3 and section 4. Mock observations of the simulated galaxies, obtained using the software
SUNRISE$^{137}$, are shown in optical wavelengths I-band (top panels)
corresponding to a particular band seen by the telescope of the Sloan Digital Sky Survey (SDSS))
and in ultraviolet bands (as seen by the GALEX ultraviolet satellite).}
\label{fig:zoom}
\end{figure}

Supernovae explosions are a likely driver of ISM turbulence since the blast-waves generated by them can transport
energy and momentum to scales as large as several hundred parsecs, perhaps up to kiloparsecs. This gives rise to dramatic
outflows of gas above the disk plane in galaxies that are actively forming stars (Figure 10). 
Other drivers of turbulence in the ISM are probably operating, both at small scales, for example
protostellar outflows, and at large scales, such as spiral waves generated by the large scale gravitational instability
of the galactic disk at scales of 1 kpc and above. Magnetic fields probably also play a role since they can generate
turbulence via magneto-hydrodynamical (MHD) waves$^{62}$.
This brief summary highlights the complexity of the ISM physics.
Things are complicated even further by the fact that the main mechanism of molecular cloud formation is 
still unclear, and so are the details of how molecular cores collapse into stars.

\begin{figure}
\centering{
\resizebox{12cm}{!}{\includegraphics{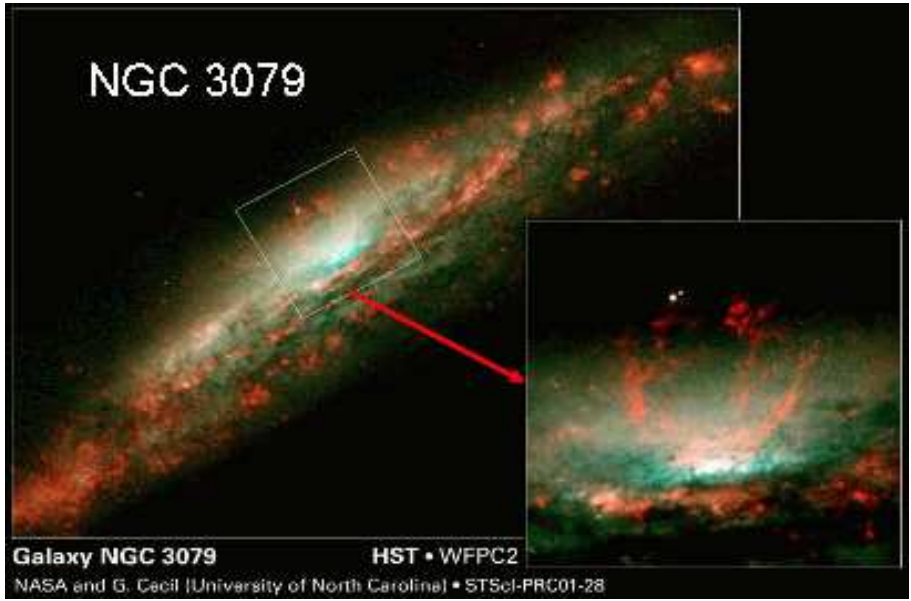}}\\
\caption{
The disk galaxy NGC3079  which shows clear evidence of outflows triggered by supernovae explosions in the 
central region.
The image was obtained with the Hubble Space Telescope (HST) (courtesy of NASA).
The gas in the outflow is heated to large temperatures and shows up in X-rays.
The central bubble is more than 3,000 light-years wide and rises 3,500 light-years above the galaxy's disk. The smaller photo 
at right is a close-up view of the bubble.
}}
\label{fig:feedback}
\end{figure}

\subsection{Modeling issues}

How can we model the complex phenomenology of the ISM in large scale cosmological galaxy formation simulations? 
This has always been a major problem
because the multi-phase structure of the ISM is also multi-scale, going from kpc scales, where most of the gas 
is ionized, to scales of tens of parsecs and below, where most of the gas is molecular, with the density changing
by about 8 orders of magnitude across the different phases.
Cosmological
simulations  follow the formation of galaxies in a volume of at least several tens of Mpc because, as
we explained in section 2.4, the generation of angular momentum and other aspects
of structure formation require that the
large scale gravitational field is properly modeled. This calls for resolving seven orders of magnitude
in length scales, and about ten orders of magnitude in density, going from tens of Mpc scales and densities of
$10^{-7}$ atoms/cm$^3$ to parsec scales and densities  $> 100$  atoms/cm$^3$.
This can only get worse if we aim at following directly the process of star formation
; indeed stars form from the collapse of the densest region of molecular clouds, 
the so called cores, at scales of $< 0.1$ pc$^{62}$. Currently, limitations in computer power as well as in the efficiency
of available algorithms, make it possible to resolve at best scales of 100 pc in
cosmological simulations (very recent simulations have been able to achieve a resolution of less than $50$ pc,
but they can only cover the first few billion years of evolution$^{43}$).
Simulations with resolution adequate to follow directly molecular clouds, interstellar
turbulence and star formation do exist, but are restricted to studying an isolated galaxy$^{64,65}$ or a small region of a single galaxy 
$^{66}$. Detailed numerical models 
of the effects of supernovae explosions also exist, but again they are restricted to a small volume of the ISM$^{67,68}$.
For this reasons, the past decade or so has seen a lot of research activity being
focused on designing the so called "sub-grid" models for simulations. These models essentially contain a phenomenological 
description of the processes occurring below the smallest scale resolved in the simulation. The phenomenological model is
incorporated into the same parallel codes that compute gravity and hydrodynamics as well as
radiative heating and cooling. 
Sub-grid models, being phenomenological, inevitably contain some
free parameters that are tuned to reproduce important observables such as the typical star formation rate
for a galaxy of a given mass, namely how much gas is turned into stars over a given amount of time.

The description of star formation is fully sub-grid, while the thermodynamics of the ISM is partially sub-grid.
What  do we mean by "partially" subgrid? One example is the following; radiative cooling is directly 
modeled for the range of densities
accessible to the simulations while its effect below the grid is only implicitly accounted for
in a phenomenological way. Since simulations typically lack resolution below a few hundred parsecs this
sets a maximum density that the simulation can resolve of order 0.1 atoms/cm$^3$, which is close to the
density of the WNM ($T \sim 10^4$ K). For this reason cooling processes that are important below $T \sim
10^4$ K are usually neglected. Likewise, radiative heating is partially determined by the thermal and turbulent 
energy injection from supernovae explosions and/or from accreting massive black holes at the center of galaxies, 
both processes being not directly resolved, but also by the ultraviolet
interstellar or intergalactic radiation background produced by stars, 
or by cosmic-ray and x-ray heating originating from various astrophysical sources, which can be directly included as constant
heating terms in the internal energy equation without the need of a sub-grid model.
Finally, interstellar turbulence cannot be resolved in galaxy formation simulations, nor it is accounted for in the sub-grid models. 
In what follows we  will recall the main features of the sub-grid models widely used in galaxy formation
simulations, pointing out their
differences. First  we will cover the star formation recipes and then sub-grid thermodynamics.

\subsection{Star formation recipes}

Star formation models used by different groups are very similar in essence. They describe the conversion
of gas into stars$^{69,70}$
A given star particle, once created, will represent a star cluster rather
than a single star because of the limited mass resolution in the simulations. Sub-grid models in this
context are all based on the idea
that stars can form only when the gas climbs above some threshold density$^{69,70}$.
This density is usually taken 
to be $0.1$ atoms/cm$^3$. Indeed this is the typical density of the WNM, and is 6-7 orders of magnitude
lower than the density of dense, star forming molecular cloud cores despite being
a relatively high density in the context of simulations that model a large cosmological volume.
Once the density of a gas parcel is above the latter threshold stars are formed
over a timescale comparable to the local dynamical time, as expected from the fact that stars emerge
from the gravitational collapse of dense gas$^{69,70}$ 
The amount of stars 
formed over such a time span is regulated by an efficiency parameter that can be tuned to match
observed star formation rates. This efficiency parameter crudely mimics the effect of
unresolved physical processes that regulate the conversion of gas into stars. Among the latter,
the formation and destruction rates of molecular hydrogen that determines how much gas
is available in the star-forming dense molecular phase, and the effect of turbulence, magnetic fields, protostellar
outflows and all those processes that determine how efficient is star formation within single clouds.
Different algorithms then are distinguished by (1) how
they implement additional conditions, based on e.g. the velocity field or temperature of the gas, as pre-requisites
for star formation, and by (2) the details of the method used to convert a single gas particle into
a collisionless star particle. 

Despite various differences, all the sub-grid star formation recipes designed so far
fare quite well at reproducing the  slope of the observed correlation between the average star 
formation rate and gas density, the so called Schmidt-Kennicutt law$^{71}$.
The latter relation is global in nature since it looks at the total amount of stars formed
in an entire galaxy, and is one of the most fundamental observables that simulations use as a benchmark
for their validity. In more detail, the Schmidt-Kennicutt law states that
the density of cold neutral gas (CNM), mostly atomic hydrogen, correlates with the global star formation
rate. This is a non trivial correlation since in reality it is only the molecular phase that
is directly related to the production of stars. On the other end, somehow the molecular phase stems 
from the neutral atomic phase once this is able to achieve
a high enough density. The success of simple sub-grid recipes in reproducing the Schmidt-Kennicutt
law is thus probably related to the fact that they are all based on a threshold gas density and a
dynamical timescale$^{72}$. Both numbers are determined by how the gas density evolves with time; this in turn
is likely controlled by the large (kpc) scale gravitational instability in galactic disks, which is resolved
in the simulations.

Nevertheless, for low mass galaxies large deviations from 
the Kennicutt relation occur, and such deviations are also evident in single star forming regions of well 
studied nearby galaxies$^{73}$. In order to reproduce the latter, more complex observational
scenario high resolution simulations of galaxies  have begun to include a sub-grid description
of the molecular phase starting from the CNM instead of having to bridge all over from the WIM and WNM to 
stars$^{74,75}$. These first attempts show that a model that incorporates the formation and destruction of  molecular 
hydrogen as a function of density
and ambient temperature allows to reproduce detailed correlations between the local star formation rate and
the local density of molecular gas as traced by various molecular tracers in observations of individual
galaxies$^{73}$.
These recent models produce realistic results for the star formation
properties  of galaxies of varying masses$^{76,75}$, despite the fact that
heating by supernovae explosions is neglected. 
They provide a way to follow
the molecular gas fraction when the resolution is not high enough to model molecular gas
formation directly by using the gas at CNM densities
as a proxy for the amount of molecular gas$^{75,77}$.
For this procedure to be sensible the number of gas particles must of course
be high enough to allow resolving the density of the CNM ($> 50$ atoms/cm$^3$). The latest cosmological
simulations of galaxy formation are just now starting   to approach this regime.

\begin{figure}
\centering{
\resizebox{12cm}{!}{\includegraphics{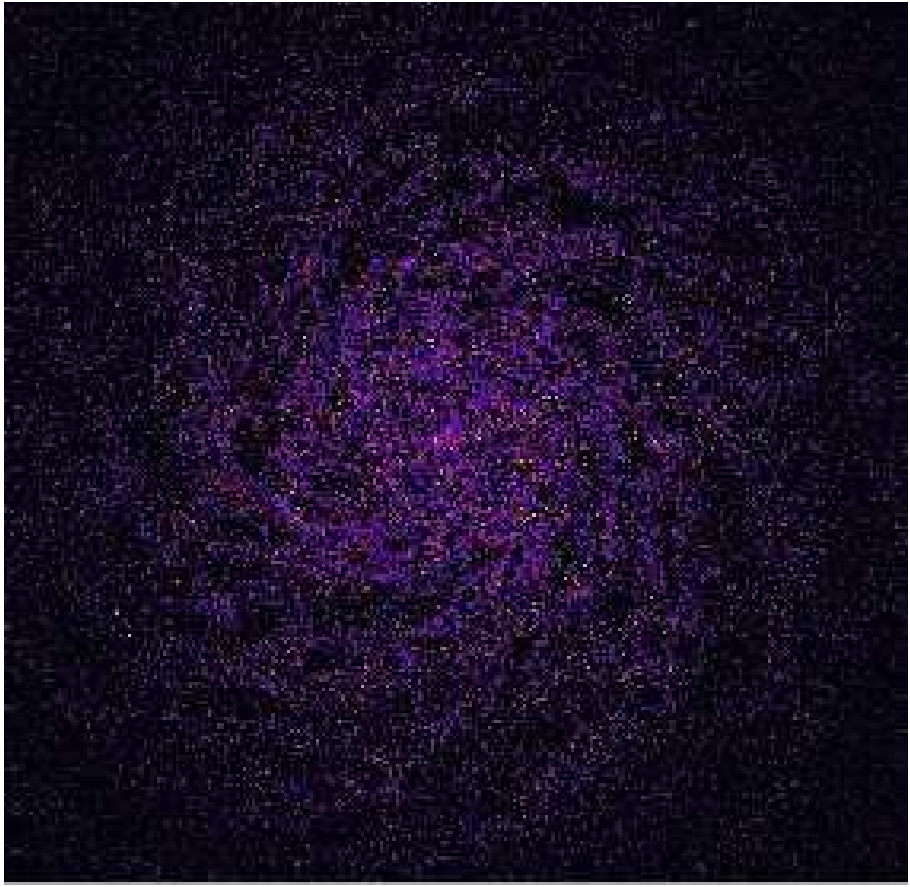}}\\
\caption{Gaseous distribution of an isolated galaxy simulated with blast-wave feedback$^{86}$ (2006 
Blackwell Publishing Ltd). Note the
holes punched by the supernovae explosions (hot bubbles) surrounded by much colder
and dense gas in filamentary structures.}}
\label{fig:feedback}
\end{figure}

\subsection{Modeling ISM thermodynamics with supernovae feedback}

\paragraph{Thermal and kinetic feedback}

Sub-grid models of supernovae feedback come in a wide variety. Early attempts in the 90s were essentially
based on two types of implementations. In the first implementation$^{69, 34}$ a fraction of the energy of individual supernovae 
explosions ($10^{51}$ erg) is damped to the surrounding gas as thermal energy. In the second method a fraction of 
the energy of the explosion is converted into kinetic energy of the surrounding gas particles rather than into
thermal energy. The latter model is motivated by the fact that the blast-wave produced as a result of the
stellar explosion will not thermalize immediately but will expand in the interstellar medium for some time as
a result of its bulk motion$^{78,79, 80}$. 
With thermal feedback the gas loses quickly the added
energy because the radiative cooling time is very short for the typical temperatures reached 
by the gas ($T\sim 10^5$ K). Eventually the kinetic energy added to the gas in
kinetic feedback is also thermalized and radiated away, but on a longer timescale. 
As a result, kinetic feedback is more effective at increasing the internal energy of the gas compared
to thermal feedback; gas cooling is counteracted more efficiently, and in cosmological simulations less gas ends up in dense
lumps. Overall the gas component loses less angular momentum while merging, resulting ultimately in larger
disks$^{79,80,81}$. However, this method 
is strongly resolution-dependent 
and is not directly informed by the actual dynamics and thermodynamics of the supernova remnant since
(the magnitude of the velocity kick given to the particles is arbitrary). 
Finally, both the thermal and the kinetic method do not account for the multi-phase structure of 
the ISM.

The rate of supernovae heating
depends on how many supernovae type I and type II go off in a given time, which in turn depends on the
adopted initial stellar mass function (IMF). The IMF, $N(m_*)$, quantifies 
how many stars of a given mass $m_*$ are present in a representative ensemble of stars.
Each star particle in the simulation is indeed not a single star but rather represents a star cluster whose unresolved 
member stars obey the chosen IMF.
Most sub-grid models assume a standard stellar mass function based on the observed
mass function in galactic star clusters. Standard mass functions are well described by power-laws,
$N(m_*) \sim {m_*}^{-\alpha}$, $\alpha < 2$, and are
dominated by stars comparable or less massive than the Sun (for a review see$^{82}$).
Recently, some researchers have
begun to distinguish between a late epoch, close to present day, in which the IMF
 is one of the standard forms, and an early epoch, ten billions of years ago or more, in which IMF
was likely dominated by stars with masses well exceeding $10$ solar masses, called a top-heavy IMF$^{57,60}$. 
Of course in the latter case the amount of energy damped by supernovae is much larger because a higher number of massive stars,
those that rapidly explode into supernovae type II, is produced for a given star formation episode.
The stronger supernovae heating rate resulting from a top-heavy IMF has a positive effect on suppressing
the formation of cold and dense gas clumps in dark halos$^{57,60}$.

\paragraph{Adiabatic feedback and blast-wave feedback}

In a third, alternative approach to feedback
the energy of the explosions is damped to the gas as thermal but the radiative cooling of such gas is stopped for a 
timescale of about $20-30$ Myrs, based on the assumption that it will take this much time for the (unresolved) interstellar turbulence 
generated by supernovae explosions to be dissipated$^{83,84}$.
Of course this is a very simplistic way of accounting for
the non-thermal energy budget of the ISM. In fact it is known that turbulence does not just
behave as an effective pressure support for the ISM but can also drive cloud collapse, thus
aiding star formation, by creating local density enhancements$^{85}$.
When applied to cosmological simulations, this "adiabatic" feedback proved even more effective than kinetic
feedback in producing an extended disk component$^{84}$. Recently, we have 
developed further this idea$^{86}$; in the remainder we will refer to our model as the
"blast-wave" model. 
In the latter model the timescale during which cooling is shut off is
self-consistently calculated based on a sub-grid model of the blast-wave produced by a
supernova explosion.
By temporarily preventing the cooling of the hot phase created by supernovae feedback this type of methods
naturally produces a two-phase medium with hot bubbles triggered by supernovae explosions ($T > 10^5$ K) 
surrounded by a colder, filamentary phase ($T \sim 10^4$ K) (see Figure 11). Based on the density and temperature range,
these two phases roughly represent the hot intercloud medium and a combination of the WNM and WIM.

The blast-wave model stems from a classic model developed in the late seventies$^{87}$ which 
reproduces the main features of the multi-phase ISM in our Galaxy.
In the latter the blast-wave sweeping the ISM undergoes an adiabatic expansion
phase during which radiative cooling is negligible. The radius of the blast-wave as a function of time can  be 
directly computed from the local physical parameters of the ISM.
In the numerical implementation such radius defines the size of the volume of gas particles that are unable
to cool during the adiabatic phase. The adiabatic phase lasts of order 30 Myrs, after which radiative losses would become
efficient and the gas is again allowed to cool radiatively. The blast-wave can affect volumes with
length scales  of up to a kiloparsec, as observed in star supernova-driven outflows (Figure 10)
; the latter are well resolved in the highest resolution cosmological simulations
available today$^{55}$.
 The delay introduced in the cooling time has also a direct
effect on star formation since the gas becomes hotter than the typical temperature of gas eligible
to form stars. The latter is a free parameter in star formation recipes; it is typically set close to $10^4$ K, the lower limit of the radiative cooling function for atomic gas comprising only hydrogen and helium (the standard composition adopted in cosmological
simulations). This
somewhat mimics the fact that the interstellar turbulence fed by supernovae explosions 
might temporarily suppress the collapse of dense molecular cores into stars.
The blast-wave feedback model has only one free parameter, the efficiency of supernovae feedback,
namely what fraction of the energy generated by the supernovae explosions is damped to the gas$^{86}$.
An individual particle belongs to one phase only and there is no
enforcement of pressure equilibrium between different phases.
Indeed in the interstellar medium pressure equilibrium applies only to the warm and cold diffuse atomic phase but
not to the star forming cold molecular phase or to the hot intercloud medium produced by supernovae
explosions$^{62}$.

\paragraph{ISM Models with effective equation of state}

Another method is based on treating the interstellar medium
via an effective equation of state that accounts for the relative contributions of a hot and a cold 
phase in a statistical fashion$^{88}$. This method is the modern version of earlier
attempts made to model interstellar gas as a two-phase medium with a hot and a cold phase$^{89}$.
Such earlier methods were splitting gas particles in two sets, cold and hot particles, with the transition 
from one to the other phase being regulated via radiative cooling and heating by supernovae feedback. Some 
of these methods were decoupling the hot and cold phase by solving the hydrodynamical equations separately for 
two sets of particles$^{51,90}$. In the new method 
gas particles come only in one species; each gas particle in the simulation is effectively
representing a finite volume of the ISM with its share of cold and hot phase being determined by the
local density$^{88}$.
The share of hot and cold phase determines how compressible is the gas, namely the
form of the function $P(\rho)$ ($P$ is the pressure and $\rho$ the density of the ISM).
The larger the fraction of hot
phase associated with a given particle the stiffer its equation of state, and therefore the higher
will be the gas pressure assigned to it for a given density. The hot phase, and thus the 
pressurization of the
gas, is the result of supernova feedback and its effects on the interstellar medium. These effects
are treated in a phenomenological way and involve the creation and destruction rates of unresolved 
star forming molecular clouds. The phenomenological equations contain free parameters.
In denser regions the equation of state becomes increasingly stiffer as star formation generates 
heating via supernovae explosions and augments the fraction of hot phase. When the hot phase increases,
however, cooling also becomes more efficient and tends to replenish the cold phase. For reasonable
values of the free parameters the simulated galaxies tend to reach quickly self-regulation, in 
the sense that the equation of state approaches a steady-state regime.
A more sophisticated version of this model also includes the additional heating resulting from
accretion onto massive black holes at the center of galaxies$^{91}$, which
typically renders the equation of state stiffer. On the contrary, switching off heating by supernovae
feedback produces a softer equation of state that approaches the isothermal regime (as expected based
on the efficient cooling rate at the density and temperatures typical of galactic disks).
Since the description of the interstellar medium given by the effective equation of state is statistical
in nature the two phases are implicitly assumed to be in pressure equilibrium locally, which is
not the case in the blast-wave feedback model. This is a major difference between the models, and
its consequences should be systematically explored in the future.
This method produces very smooth disks in the simulations when feedback is included$^{56}$ 
(without feedback the isothermal disks undergo rapid fragmentation owing to gravitational instability) 
as opposed to disks rich of filamentary structures and flocculent spiral arms arising in 
the blast-wave model (Figure 11).

Recently, the effective equation of state model was modified to allow for 
strong supernovae feedback due to
a top-heavy stellar mass function arising in specific conditions$^{57}$, such as during galaxy mergers (there are indications
that the high star formation rates produced by mergers might lead to a top-heavy stellar mass function by
changing the balance between heating and cooling that controls the collapse of molecular cores$^{92}$).
This produces
a "burst mode" in which heating is so efficient that it can lead to partial evaporation of the ISM
in low mass galaxies. This appears to reduce the mass of the bulge and make a galaxy more
disk dominated, although galaxies always end up with {\it some} bulge component.

\section{The effect of ISM models on disk formation}

\subsection{Isolated galaxy models}

Testing the effect of sub-grid models in galaxy formation simulations   
is best done not in cosmological simulations but in high-resolution numerical realizations of
isolated galaxies$^{93,46}$ (see section 2).
These models are constructed on purpose to resemble observed galaxies, with a disk of gas 
and stars, eventually
a stellar bulge, and always an extended, massive dark matter halo with structural properties (mass, spin,
density profile) consistent with the results of cosmological simulations describing the hierarchical growth  
of CDM halos. The various components are in dynamical equilibrium, representing a stationary
solution of the relevant dynamical equations for collisionless and fluid matter. Small deviations
from equilibrium are present as a result of discretization and other approximations introduced by
the numerical methods$^{93}$. These models by-pass all the issues of ab-initio galaxy formation
and are built on the assumption of angular momentum conservation during the baryonic collapse.

Tests of the blast-wave feedback model$^{86}$ using such equilibrium galaxy models show that
after a while the star formation rate and phase diagram of the galaxy become nearly steady-state 
(self-regulation). Locally, however, the phases are not in pressure 
equilibrium and a patchy, filamentary structure with bubbles arises that is well resemblant of observed 
galaxies (Figure 11). Other important properties of observed galaxies, such as the volumetric ratio between
cold and hot gas, the thickness of the gaseous disk and the ratio between ordered, rotational motions
and random motions in the stellar component, are also satisfactorily reproduced$^{86}$.


\begin{figure}
\centering{
\resizebox{6.5cm}{!}{\includegraphics{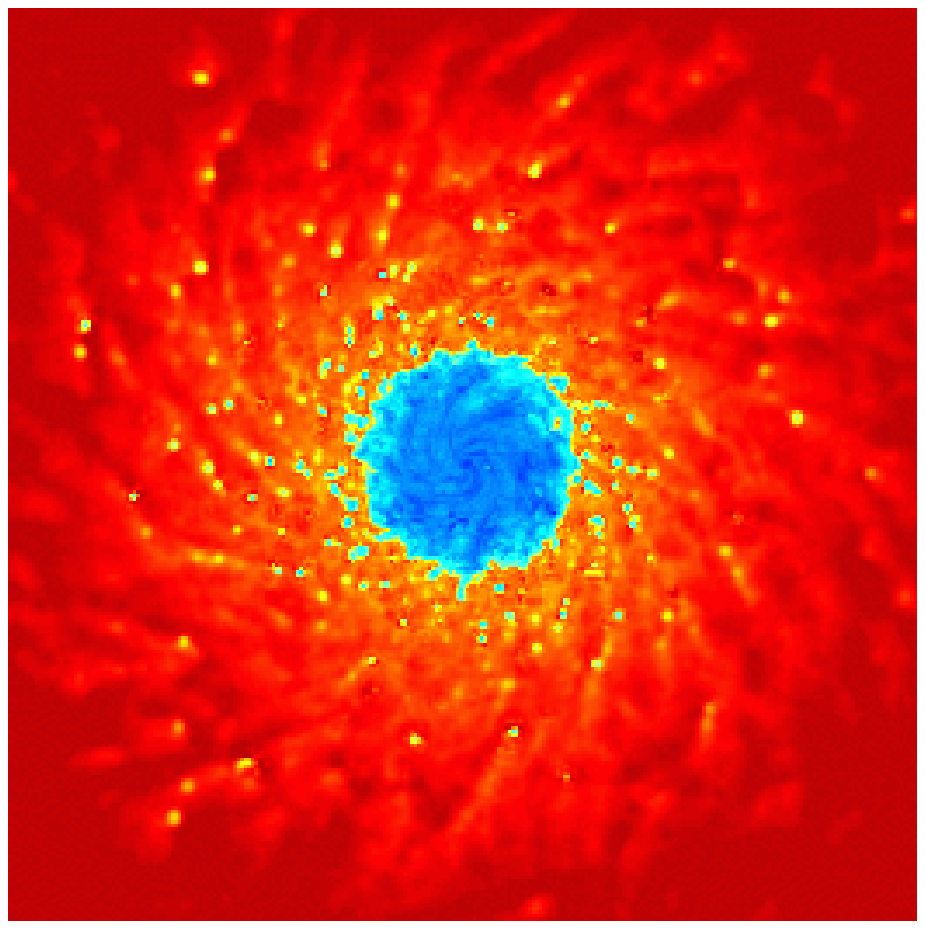}}\\
\resizebox{6.5cm}{!}{\includegraphics{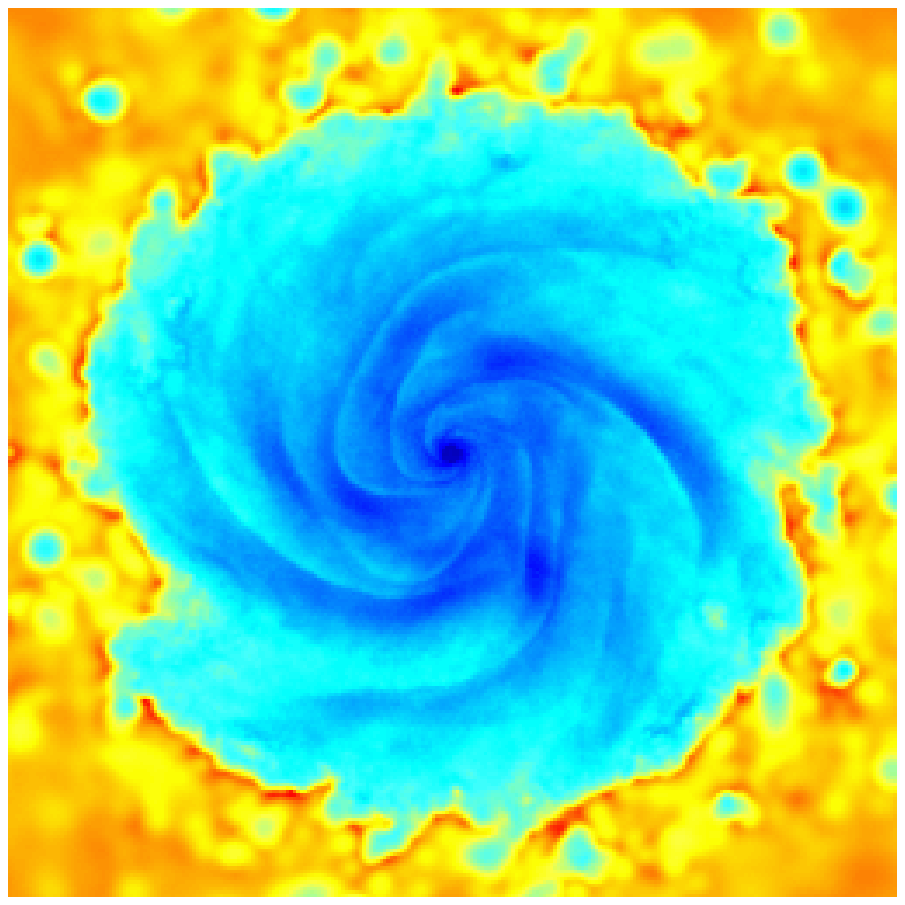}}\\
\resizebox{6.5cm}{!}{\includegraphics{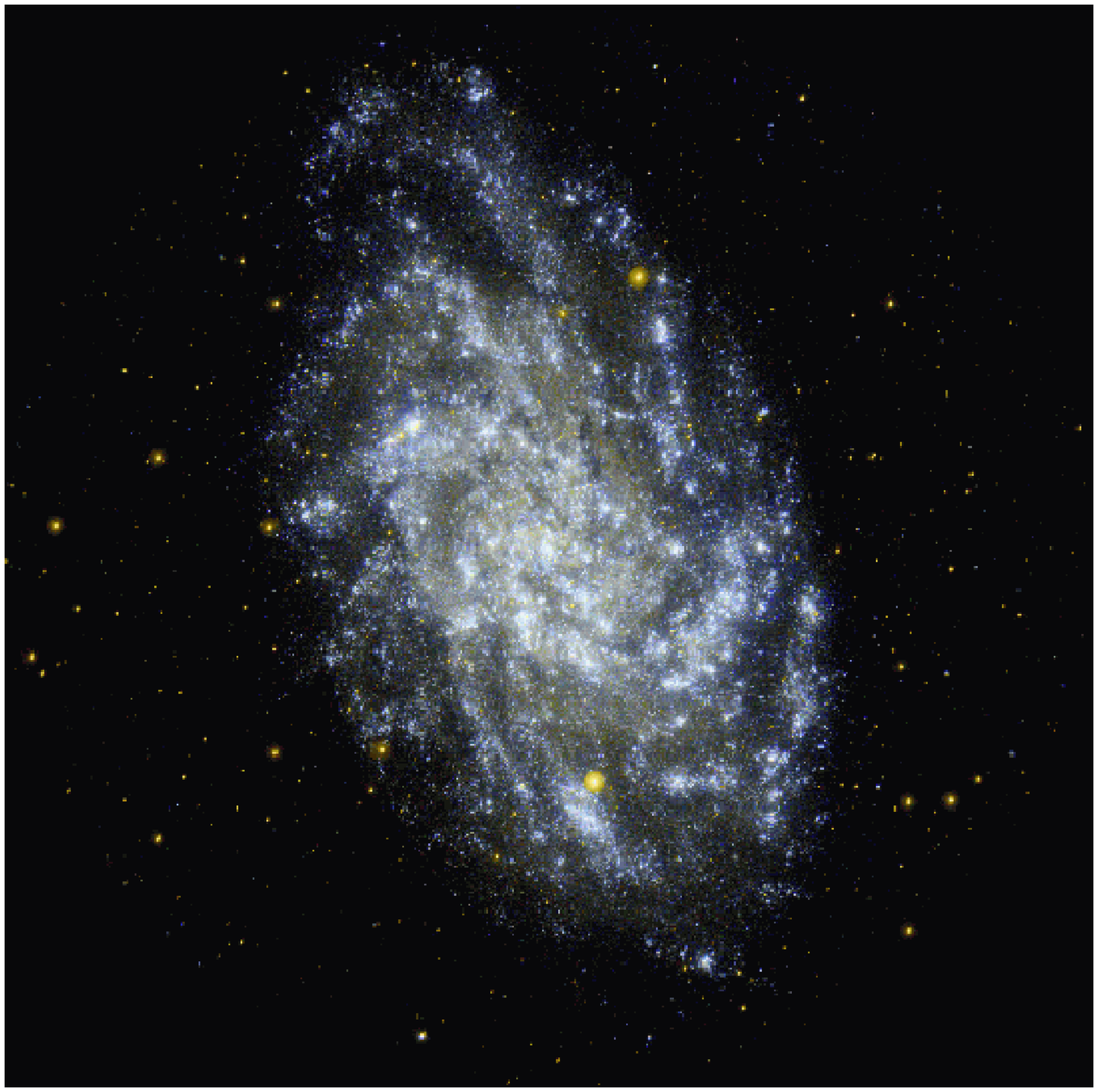}}\\
\caption{Top and middle panel: snapshots of a high-resolution galaxy formation simulation in an 
isolated halo that produce a bulgeless galaxy thanks to a low baryon fraction$^{48,130}$ (2006, 2007 
Blackwell Publishing Ltd). See section 4 for details. The top panel 
shows a region of 60 kpc on a side in which the cold clouds produced by thermal instability in the infalling gas
are visible. The middle panel zooms in the inner 20 kpc, showing the disk-dominated galaxy with
flocculent spiral arms that resembles the bulgeless nearby M33 galaxy. Bottom: the nearby bulgeless 
galaxy M33.}}
\label{fig:feedback}
\end{figure}

These idealized numerical experiments are useful for validating
the phenomenological recipe contained in a given sub-grid model. Nonetheless, they give limited 
information on what 
happens in a realistic cosmological setting in which galaxies are always out of equilibrium
as a result of mass accretion from their surroundings and frequent
merging events with other galaxies. 

An intermediate level of realism between equilibrium galaxy models and a fully
cosmological simulation is represented by models that follow the formation of a single galaxy from
the cooling and collapse of gas within an isolated dark matter halo. These models are more realistic
because they do not start from an idealized equilibrium condition and because they do follow the
assembly of the galaxy. However they differ from cosmological simulations
because they neglect the external tidal field and the merging with other halos/galaxies during the
galaxy assembly process$^{48,50}$.
One fixes the specific angular momentum of the gas by choosing a value of the spin
parameter. Once the spin parameter is fixed, for example to the mean value found for dark halos 
in cosmological simulations, $\lambda \sim 0.035$,
the baryonic mass fraction, namely
gas mass initially present in the dark halo,  will control the properties of the galaxy that
will emerge from the collapse.
If the baryonic mass fraction $f_b$ ($f_b=M_{baryon}/M_{vir}$, where $M_{baryon}$ is the mass of baryons
in the halo and $M_{vir}$ the total halo (virial) mass, i.e. dark matter + baryons)
 is fairly high, comparable to the universal baryon
fraction expected in the standard LCDM cosmology ($f_b \sim  0.17$), a massive disk forms and 
undergoes soon a bar
instability, this being a  common outcome of gravitational instability in a rotating gaseous or stellar disk.
As we mentioned in section 2.4, the bar transfers angular momentum outward and mass inward, leading to a steep 
stellar density profile in the center and a much shallower profile in the outer part.
When a standard star formation recipe is included in the simulation the bar becomes mostly stellar 
because gas is efficiently converted into stars, and undergoes another form of gravitational instability in the 
vertical direction, known as "buckling instability"$^{94}$. The latter instability
deflects the orbits of stars away from the disk plane, turning the elongated bar into a 
roundish, bulge-like component.  The latter is widely recognized as an important mechanism
of bulge formation, alternative to mergers between galaxies$^{95,96,54}$.
After the buckling instability the simulated 
galaxy looks like our Milky Way or the Andromeda galaxy, with a bulge surrounded by a more massive 
and extended disk of stars and gas$^{48}$ (Figure 4).

\begin{figure}
\centering{
\resizebox{12cm}{!}{\includegraphics{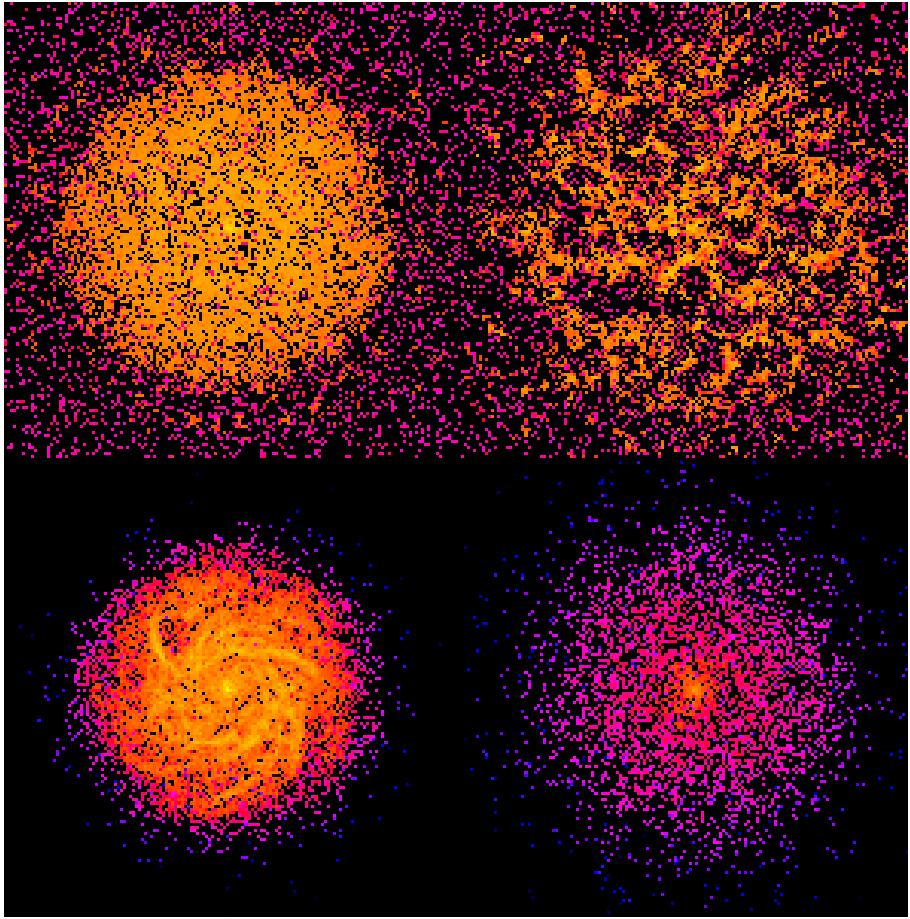}}\\
\caption{
Comparison of disk galaxy (isolated collapse simulation) with thermal
feedback$^{48}$ (left, a temperature floor is used - see text) and blast-wave feedback$^{86}$ (right).
The gaseous disk is shown on top and the stellar disk is shown at the bottom. Boxes are 20 kpc
on a side, the galaxy halo has $V_c = 70$ km/s before collapse, simulations have $5 \times
10^5$ SPH particles and $10^6$ dark matter particles and are shown after about 1 Gyr.
}}
\label{fig:feedback}
\end{figure}

If the fraction of baryonic material is chosen low enough it is possible to reproduce a disk galaxy without 
a bulge component. The bar does not form, and thus no spheroid is produced, because this time the disk
has a lower mass and can hardly become gravitationally unstable. 
The final galaxy looks similar to our neighbor bulgeless galaxy M33 (Figure 12).

\paragraph{The origin of exponential stellar density profiles}

The M33-like galaxy emerging from the simulation with a low baryon fraction does not exhibit an exponential 
stellar surface density profile such as that of typical bulgeless galaxies, rather it shows a sharp upturn 
within a few hundred parsecs from the center. Indeed, this
upturn is caused by a dense clump of gas and stars, containing of order $10^9$ solar masses. Central
stellar concentrations of comparable masses, known as stellar nuclei, are seen in some bulgeless galaxies but 
their typical sizes are ten times smaller, so that they do not affect the surface density profile at scales 
of hundreds of parsecs$^{97}$.
Moreover, a large fraction of bulgeless galaxies does not have such nuclei and shows 
a stellar exponential profile across several kiloparsecs in radius. This is currently one of the
most important problems of disk formation.
The location of a parcel of gas in the disk after the collapse, namely how close to the center
such a  parcel settles into centrifugal equilibrium, determines the gas density profile,
which in turn sets the stellar density profile via the conversion of gas into stars. 
Such location is determined by the combined action of 
gravity, pressure and rotational support. The presence of a dense concentration in the simulations might
suggest a "second" angular momentum problem at small scales. It is a second problem because it occurs
even when the  total spurious angular momentum loss
in the disk is down to a few percent thanks to high resolution$^{48}$. 
As we mentioned in section 2, numerical loss of angular momentum near the 
center of the potential is hard to tackle even at high resolution, and might be playing a role in this
case.
However, this cannot be the only issue. In fact, one dimensional spherical numerical models of
gas collapse in cuspy CDM halos that impose angular momentum conservation also predict
that same inner upturn of the stellar density profile$^{10}$. 
This seems to be a result of how the gas
settles in centrifugal equilibrium in the cuspy potential of a CDM halo$^{10,13}$. The upturn in 
the profile is caused
by the collapse of the innermost shells of the gas distribution, and these are the ones that
collapse first because they have the shortest free-fall time. 

One possibility to
reconcile the simulations with the observations would be to alter the halo potential -- if the halo has
finite density core its potential well would be shallower and the gas would likely satisfy the 
the centrifugal equilibrium condition
further out from the center, avoiding the formation of the dense central clump. This of course would
require to invoke some mechanism that modifies the halo profiles predicted by the CDM simulations, or even to
resort to an alternative cosmological model. Both options should not be disregarded but one
may wonder whether such a drastic change of scenario is really required.
Since pressure also
plays a role in deciding the radius at which gas comes into centrifugal equilibrium, as well as its distance
above and below the plane, hence ultimately its mass distribution, the answer could lie in the details of gas 
thermodynamics during disk assembly.

Here we present preliminary results suggesting that the model assumed
for energy feedback in the disk has a big impact on the slope of the stellar density profile.
We recomputed the same disk formation models presented in already published work$^{48}$ using the
blast-wave feedback model (40\% of the energy of the supernovae explosions is damped to the
gas in this particular simulation). The gas disk has an irregular
and flocculent structure without a significant central concentration, at odds with
the simulation without feedback (Figure 13).
The stellar disk has a lower average density and its surface density profile 
has a much flatter slope compared to the case without feedback (Figure 14)  and shows a milder
upturn at small scales; what is most important, however, is that such upturn appears only after 
a couple of Gyrs of evolution. Therefore
it is related to the internal evolution of the disk (spurious and/or physical) rather than being
caused by the initial conditions (i.e. by gas collapse in a cuspy halo profile).

\begin{figure}
\vskip 10truecm
{\includegraphics{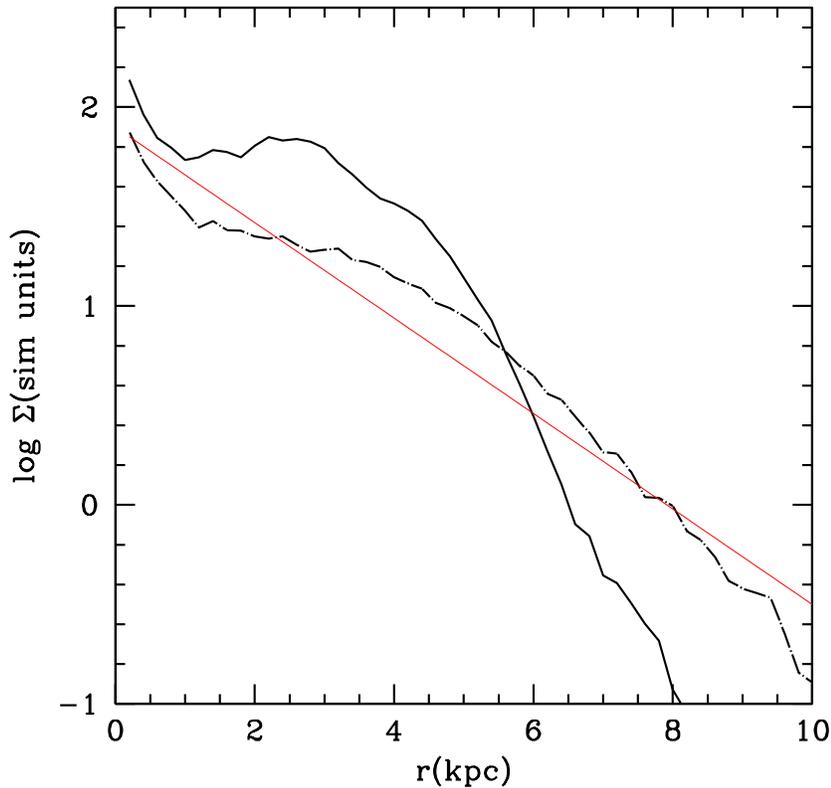}}
\caption[]{\small 
Stellar surface density profile (in code units) of a galactic disk in the blast-wave collapse simulation
(dot-dashed line) and in the collapse simulation with a temperature floor mimicking thermal feedback 
(solid line). The red line shows a possible exponential fit to the surface density profile
to highlight the different slopes.The profiles are shown after about 2 Gyrs of evolution.}
\label{fig:zoom}
\end{figure}

A single exponential curve can reproduce the slope of most of the stellar
mass distribution, which clearly was not the case with thermal feedback (see Figure 14).
So now the problem has shifted from forming a pure exponential profile to
preserving it for a timescale long enough, i.e. many Gyrs, to explain
why we observe many pure exponential disks in the local Universe.
The fact that the stellar profile becomes closer to exponential with using blast-wave
feedback is not trivially related to the higher pressure support in the gas disk produced by the
supernovae explosions.  Indeed the simulations that we previously published$^{48}$
included a minimum temperature below which the gas was not allowed to cool in order  to crudely mimic the heating due to feedback.
Such temperature floor was comparable to the mean temperature of the gas in the blast-wave feedback simulation
($30000-50000$ K). 
After all, the models with a temperature floor can explain observed trends of increasing thickness of the disk, 
lowered star formation efficiency and higher gas fractions in the disk  with decreasing galaxy mass 
due to the enhanced pressure support$^{98}$.

However, while the temperature floor is by construction the same  everywhere in the disk, the blast-wave 
feedback produces a two-phase medium such that density, pressure and temperature can have 
significant local variations (see previous section).
Indeed the morphological appearances of the two disks are very different (see Figure 13), with the blast-wave
model producing a  much more flocculent and irregular gas disk which
resembles more closely that of late-type low mass galaxies such as the nearby Large Magellanic Cloud$^{86}$.
The bubbles produced by supernovae literally punch holes in the disk and push the neighboring 
cold gas squeezing it into dense filaments, where star formation can go on. The fact that the cold gas is confined to
filaments and cannot fill a large volume naturally avoids the formation of large dense clumps of gas, possibly explaining
why the blast-wave feedback is more effective at suppressing the upturn of the surface density profile compared
to a model with a uniform temperature floor.

Cosmological simulations adopting the model of the effective equation of state$^{87,91}$ were also
able to produce a disk galaxy having a nearly exponential profile without a central dense clump$^{56}$
but the stellar disk was too thick and was rotating too slowly in the central few kiloparsecs compared 
to typical disk-dominated galaxies  - the inner disk was deficient in specific angular momentum by a factor of 2.
This suggests
that the pressurization of the interstellar medium produced by this model, which determines the vertical structure
of the disk, might be excessive and might disfavour the formation of a realistic thin disk.
These results reinforce the idea that the solution of the exponential disk
profiles lies in a correct model of the thermodynamics of the interstellar medium and star formation
rather than in drastic modifications of the underlying cosmological model.

\subsection{Cosmological simulations of disk formation with blast-wave feedback}

When applied to galaxies of a range of masses forming in a cosmological simulation
the blast-wave feedback model allows to improve significantly the match with real galaxies 
in at least three ways compared to the case in which a simple thermal feedback model is used$^{55,100}$:

\bigskip
(1) At a given resolution it produces more extended disks with a smaller bulge (although  increasing 
the mass resolution has the strongest effect in reducing the bulge-to-disk mass ratio), in line 
with what was found for isolated collapse experiments (Figure 9).

\bigskip

(2) It produces automatically the right trend of star formation histories with galaxy mass (Figure 15).

\bigskip

(3) It produces galaxies that lie close to the observed correlation between rotational velocity
of the disk and luminosity of the galaxy, also known as Tully-Fisher relation$^{99}$.

\bigskip
(4) it allows to predict correctly the  stellar mass -metallicity relation$^{100}$ measured in local galaxies$^{101}$ 
and at $z=2$$^{102}$, according to which more massive galaxies are also more  metal rich (Figure 15).

\bigskip

The Tully-Fisher relation
is directly connected with the most important aspects of galaxy formation$^{103, 104}$.
namely the amount of gas which forms stars  and thus determines the luminosity of the galaxy, and the depth 
of the gravitational potential well, which determines the magnitude of rotation.
In the past numerical simulations have failed to reproduce this relation. Not surprisingly, the 
only other work in which simulated galaxies were falling close to the observed Tully-Fisher is the 
same work in which an
exponential, disk-dominated galaxy was obtained$^{56}$. In the latter work the authors
also obtained a nearly flat rotation curve for one of their simulated galaxies thanks
to the absence of a dense and massive bulge. This  single rotation curve
was flatter than that of the galaxies in $^{55}$, although a new series  of blast-wave 
feedback simulations, which consider a larger variety of initial
conditions, also includes some objects with nearly flat rotation curves (Governato et al., in preparation).
A flat rotation curve with a very small bulge component is also obtained in recent AMR simulations
that, thanks to a resolution better than $50$ pc, are able to follow directly the dynamics and thermodynamics
of bubbles produced by supernovae explosions, although the calculation is carried out only until
$z=3^{43}$.

Regarding the third point, the trend between star formation history and galaxy mass is such that lower mass
galaxies have progressively more extended star formation histories, namely a larger fraction of their stars
forms at progressively later epochs. This is the observed "downsizing" of galaxies, namely
the fact that smaller galaxies appear to have formed more recently$^{105}$.
Taken at face value, for years downsizing has
been considered as a fundamental problem of the cold dark matter model, which predicts
that structure formation proceeds from small to large halos in a bottom-up fashion. The
blast-wave feedback model has allowed to resolve this discrepancy  by  
decoupling the evolutionary timescale of the baryonic component 
from that of the  dark matter component owing to their different energetics; 
in low mass galaxies gas is less gravitationally
bound to the halos and can thus be more dramatically affected by the heating due to supernovae explosions$^{106}$, with the result that star formation is quenched and more gas is available at later epochs to keep
forming stars. In other words, the star formation in small galaxies is diluted over a much longer
time span, so that when their light is measured they appear young today despite the fact that their halos assembled early.
Dilution also means a lower average star formation rate relative to large galaxies (Figure 15).
another well established observational distinction between low mass and high mass galaxies.

A lower average star formation rate also implies that at the current epoch low mass galaxies 
should be more gas rich relative to high mass galaxies. This happens in the simulations$^{55}$ and
is in agreement with the observed trend of increasing gas fraction towards decreasing galaxy mass$^{107}$.
The gas fractions in the blast-wave feedback runs are actually significantly higher than in models that use a purely thermal
feedback. 

Realistic gas fractions and star formation rates are also obtained  with
simulations employing the effective equation of state$^{56}$.
Unfortunately, a direct and systematic comparison between the latter model and the blast-wave feedback 
model is still missing. Therefore at present it is unclear
which of the two models performs best against a large set of properties measured for real disk
galaxies, especially it is unknown how the two models compare when they are used in cosmological
simulations starting with identical initial conditions. Such a comparison would certainly be of
great benefit in order to ponder in a critical way the conceptual modeling behind the two different methods.

Finally, the fact that cosmological simulations with blast-wave feedback reproduce the
observed stellar mass-metallicity relation$^{100}$ suggests that this kind of sub-grid model
not only provides a reasonable description of the star formation process but also
of the associated metal enrichment (in the model metals are liberated when supernovae
type I and type II explode, as well as via stellar winds$^{86}$) and mixing processes in galaxies.
 It is important to note that with this model of feedback galaxies
with a rotational velocity $> 50-80$ km/s do not experience major gas "blowouts" and are
able to retain a relatively large gas fraction within their virial
radius. Rather  supernovae feedback  makes star formation relatively inefficient in
small galaxies, making them gas rich and more metal poor, as the
metals produced in stellar explosions are diluted over a larger  gas
fraction. This result opens the door to the possibility of using the blast-wave 
feedback model to study the photometric  properties of very high redshift galaxies, where the
poorly known amount of metals in the gas component can strongly affect
their observational properties.

\section{Open issues and final remarks}

It is fair to say that the progress achieved by several groups in simulating galaxy formation is quite 
remarkable. Just a decade ago the spatial resolution of cosmological simulations barely reached a 
kiloparsec, which corresponds to the characteristic exponential scale length of galactic disks. Today simulations
are approaching a resolution better than a hundred parsecs, an improvement of more than an order of magnitude. 
Therefore we have moved from a situation in which the process of disk galaxy formation
was barely within the reach of cosmological calculations to a situation in which the simulations
can effectively target such process. At the same time, sub-grid models of those astrophysical mechanisms that
determine the thermodynamics of the interstellar medium, and consequently the process of star formation, 
are becoming increasingly more realistic. Furthermore, as the resolution of simulations increases further we might
enter a regime in which the multi-phase ISM and the bulk effect of supernovae explosions can be modeled
directly rather than being incorporated in a sub-grid fashion. Indeed the most recent (AMR) cosmological 
simulations at the time of writing suggest that the many assumption of sub-grid models cease to be
required if a resolution of at least $50$ pc can be achieved$^{43}$.
Finally, we have gained a much better understanding of the various numerical
effects that can dramatically affect the results of the simulations and render a comparison with observed
galaxies quite meaningless. 

Yet, despite the tremendous improvement in
numerical resolution, cosmological simulations are still affected by resolution issues in the early phase of
galaxy assembly, namely during the fist few billion years of evolution. 
The latter phase is characterized by frequent mergers (the merger rate of dark halos declines dramatically
after $z=1$, about 8 billion years ago, in the $\Lambda$CDM model) and is responsible for the build up of the
central region of galaxies, including their bulge$^{48, 55}$.
The central region of a virialized object indeed assembles earlier because it corresponds to the peak of the
local density field, which becomes gravitationally unstable and collapses earlier than the the outer, lower
density regions. The collapse of the central region is of course lumpy; many small halos, each of them
possibly already hosting a previously assembled small galaxy, come together and merge. Even in the best
cosmological simulations currently available, these "primordial" halos  are poorly resolved due to their small masses. 
Each object being resolved by only a few tens of thousand particles, spurious angular momentum loss, two-body heating and 
other numerical effects are particularly severe$^{46,47,48}$. 
This lack of resolution at early times probably explains why the same simulations that are finally producing disks with realistic structural properties $^{58, 59, 34, 55, 60}$ still exhibit central regions with an excess 
of low angular momentum material and an ubiquitous massive, dense spheroidal bulge component. Since 
bulgeless galaxies are abundant in the present-day Universe as well as several billions of years ago$^{5}$, 
here we are facing a major problem.

\begin{figure}
\centering{
\resizebox{10cm}{!}{\includegraphics{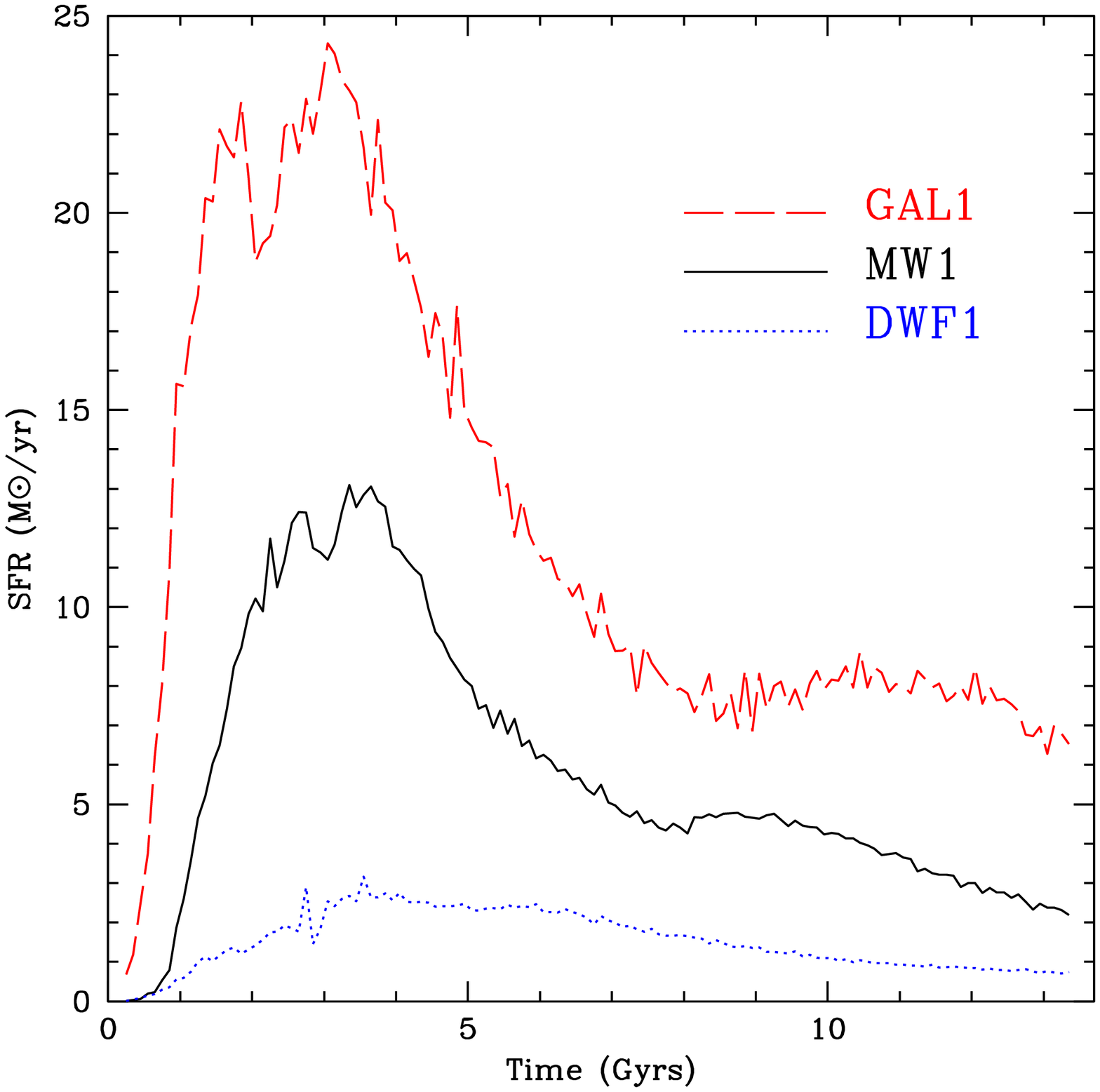}}\\
\resizebox{10cm}{!}{\includegraphics{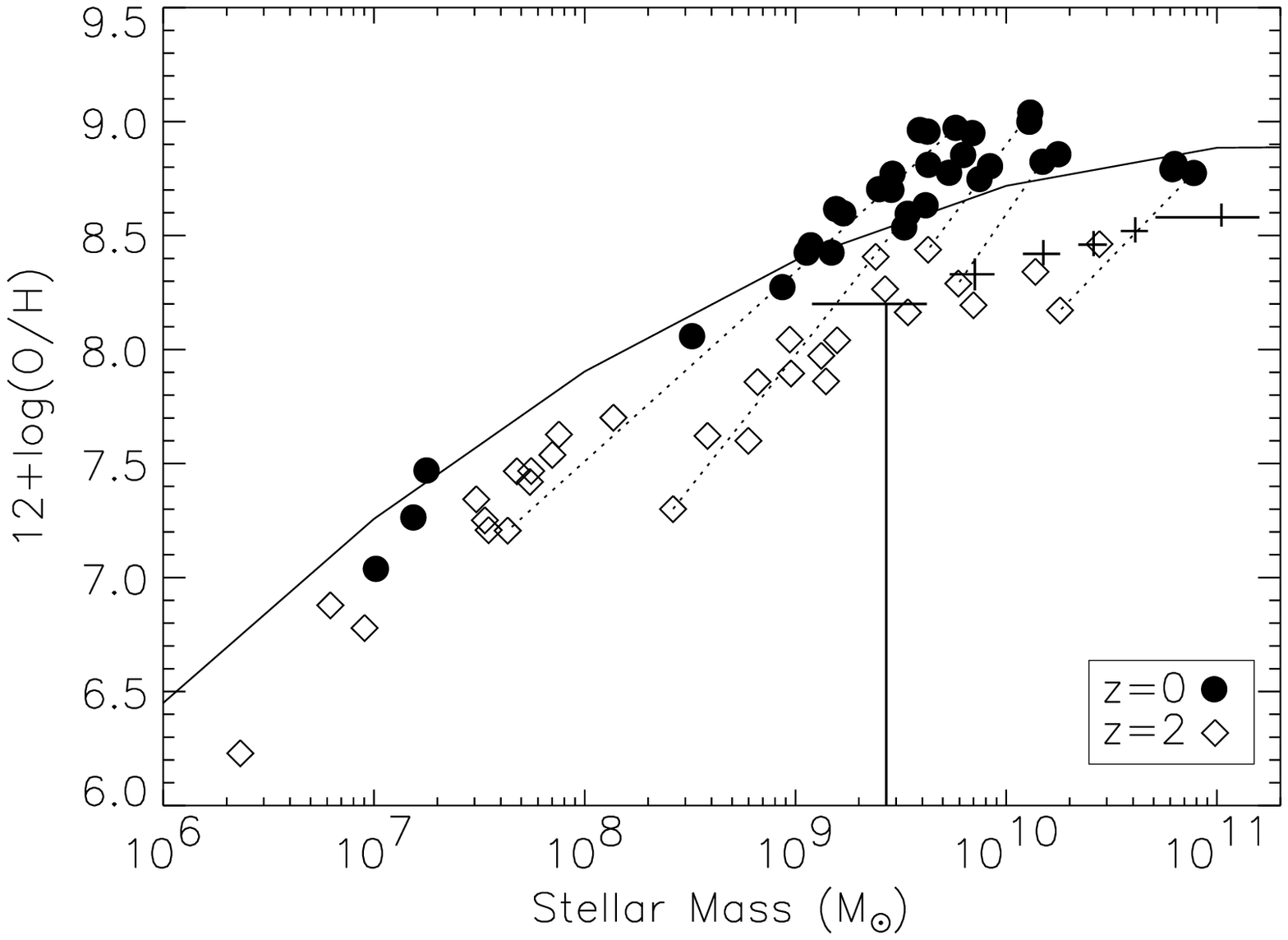}}\\
\caption{Top:Star formation histories of simulated galaxies of varying mass in high-res cosmological simulations$^{55}$ (2007 Blackwell Publishing Ltd). The
three galaxies have the following masses within the virial radius;$1.6 \times 10^{11}$ solar masses
(DWF1), $1.15 \times 10^{12}$ solar masses (MW1) and $3.1 \times 10^{12}$ solar masses (GAL1).
Bottom: Mass-metallicity relation for simulated galaxies at $z=0$ (filled circles) and $z=2$ 
(open diamonds)$^{100}$ (reproduced by permission of the AAS, 2007).
The solid curved line is the observational fit to $>53,000$ galaxies in the Sloan Digital Sky Survey$^{101}$, 
shifted down by $-0.26$ dex as found in the observations$^{102}$. Error bars show the observational mass-metallicity relation at 
$z=2$$^{102}$. 
Dotted lines connect some of the  $z=0$ galaxies to their progenitors at $z=2$ , showing 
how galaxies evolve in the  plane with time.}}
\label{fig:feedback}
\end{figure}

\subsection{Origin of bulgeless galaxies}

The ubiquitous bulge component in simulated galaxies triggers the following question;can a disk 
galaxy without a bulge ever be formed in a $\Lambda$CDM Universe?  Answering this question
using the current simulations seems a daunting task. The situation becomes even worse 
if we consider that halos with larger angular momentum, thus favourable to the formation of large, dominant
disks, are also those that are more prone to form a massive bulge because they undergo a larger number of major mergers$^{108}$.
If the main problem is resolution during the early phase of structure formation, one could hope to minimize numerical
loss of angular momentum by increasing the resolution further. However, due to the scale-free nature of collapse
in a cold dark matter Universe, when the mass resolution is increased not only previously resolved halos are sampled
by more particles, but also new, previously unresolved halos appear that are once again modeled with too few particles.
In other words, in cold dark matter simulations of any resolution there will always be some epoch at which most lumps of dark matter 
and baryons become small enough to be poorly resolved, thus biasing the angular momentum content of the final galaxy. The latter argument,
however, strictly applies only to dark matter.
As we have seen, mechanisms such as supernovae feedback can decouple the collapse
history of baryons from that of the dark matter. In addition, as the mass of the progenitor lumps decreases 
there is another important process that can lead to a significant decoupling. Between one and five billion years after
the Big Bang, namely between $z=3$ and $z=1$, both the formation rate of stars in galaxies and the mass growth of 
supermassive black holes that shine as quasars at the
center of them reach their peak, producing so much ultraviolet radiation to ionize most of the hydrogen in
the Universe$^{109,110,111, 112}$. 
This ubiquitous ultraviolet radiation, known as the photoionizing background, keeps the intergalactic
gas at a temperature exceeding $10^4$ K, suppressing the collapse of baryons in halos with
masses below $10^8$ solar masses (deeper potential wells are needed to confine gas that cannot cool
below $10^4$ K)$^{113,114,115,116}$. 
The ionizing background also causes photoevaporation of gas that has previously
collapsed in such small halos$^{117}$. The net result is that
halos with masses below $10^8$ solar masses become nearly empty of baryons during this epoch, retaining only the stars that
were formed before the rise of the ultraviolet background; when
they merge, they bring very little baryonic mass and thus should contribute little to the 
formation of bulges. 
Interestingly, galaxies with little or no bulge are mostly found among the lowest mass disk galaxies (Figure 2), which were formed by the 
hierarchical merging of smaller lumps, as expected if the photoionizing background played a key role in suppressing bulge formation early on.

Recent cosmological simulations by the major groups involved in this area of research include the effect of the photoionizing background$^{56, 58, 59, 34, 55, 60}$.
The inclusion of the ultraviolet background in the picture introduces a scale, $\sim 10^8$ solar masses, in an otherwise scale-free
structure formation model. This suggests that simulations should try to obey the following resolution requirement
in order to avoid spurious angular momentum loss during the early phase of galaxy formation; the smallest baryon-rich lumps,
namely those with masses $> 10^8$ solar masses, should be resolved  by about a million SPH
particles, and a few $10^5$ dark matter particles, as suggested by resolution studies$^{46, 48}$. The latter resolution requirement translates
into an SPH particle mass of about $10^2$ solar masses, a couple of orders of magnitude
higher than that currently achieved in cosmological simulations of galaxy formation$^{55}$.

Provided that resolution issues are solved, a conceptual problem still remains. It is a common assumption,
supported by a large number of detailed three-dimensional simulations,
that mergers between nearly equal mass disk-dominated galaxies would produce a spheroidal, 
bulge-dominated system$^{118, 119, 120}$.
These nearly equal-mass
mergers are frequent in  hierarchical assembly, especially early on. A bulgeless galaxy should only
arise if the last major merger occurs when the cosmic ultraviolet flux is still
high, at $z> 1$, i.e. more than ten billions of years ago, and involves lumps small enough to be nearly devoid of baryons.
Major mergers between more massive lumps, that were barely affected by the ultraviolet flux, and/or occurring later, when the 
flux has declined, will inevitably build a bulge;
 the bulge will only become less concentrated as the resolution increases but will not disappear.
This means that forming bulgeless galaxies
in $\Lambda$CDM needs a requirement on the merging history in addition to that on the resolution$^{121}$. 
The fact that major
mergers cease to be common after $z=2-3^{122}$ is quite encouraging
in this respect. This figure becomes even more favourable, at least qualitatively, for halos with masses below
that of the Milky Way halo, which would be consistent with the fact that most bulgeless
galaxies are of low mass (Figure 2). Yet, at the moment it is unclear 
how these figures on halo merger rates can be compared quantitatively with the fraction of bulgeless galaxies seen 
both today and at $z=1$ in large galaxy surveys such
as COSMOS$^5$.

\smallskip

The final structure of the galaxy is strongly dependent on its merging history not only because
the latter contributes to determine its bulge-to-disk ratio but also because it affects the
structure of the disk itself$^{123}$.
After the last major merger the galaxy grows via
accretion of smaller lumps of dark matter and baryons (usually referred to as satellites) 
and gas accretion from the hot gas cooling from the halo.
We could name this second phase "oligarchic growth", using a terminology
well known in the field of planet formation, where a similar switch between modes of growth
occurs in the case of colliding km-sized rocky bodies (planetesimals). 
Bulgeless galaxies should arise in those systems that switch to "oligarchic growth" earlier than
the rest of the galaxy population based on our previous argument.
However this simple prediction is complicated by the fact that a bulge could also arise during the 
"oligarchic growth" phase as a result of bar formation and subsequent buckling$^{54}$.
Therefore, one should require that the fraction
of $\Lambda$CDM halos that switch to "oligarchic growth" at high redshift be comfortably larger
than the observed fraction of bulgeless disk galaxies since some of them might have
formed their bulge later via internal evolution.

It is important to consider the possibility that the accepted assumption that mergers produce bulge-like,
spheroidal components might not be true in general. Recently, it has been shown that if
the disks of two galaxies contain a lot of gas a large disk can be produced as a result
of the merger$^{124,125}$
This is because the gas dissipates and settles 
into centrifugal equilibrium at fairly large radii owing to the angular momentum originally locked
in the orbital motion of the two galaxies.
The disk re-growth might actually increase
the disk-to-bulge ratio provided that there is sufficient gas in the original galaxies,
but this still needs to be quantified.
Now,  interestingly galaxies appear to become more gas rich at earlier
times and for lower masses. It is thus plausible that at early times galaxies underwent
several gas-rich mergers that contributed to building a disk
more than to forming a bulge. 
As a consequence, it seems relevant to carry out simulations that are very accurate in the
early phase of galaxy assembly even if this prevents from following the formation of the galaxy until the
present epoch (Callegari, Mayer et al., in preparation).

\smallskip

As we have seen in the previous section,  mergers are not the only way by which a central mass concentration
can arise in galaxies. The non-cosmological simulations also exhibit a profile that becomes
steeper than exponential towards the center, although models with blast-wave feedback 
look very promising as fas as limiting the growth of the central density enhancement.
We note, however, that the initial gas density and angular momentum distributions of the isolated models are poorly constrained.
Testing  different initial conditions can teach us about the physical conditions of the gaseous halo needed in order to 
be consistent with observations. Preliminary simulations (Kaufmann et al. in preparation) show that an initial 
gas density profile with a core, and thus a high initial entropy (as expected in scenarios in which some 
non specified primordial heating mechanism, perhaps feedback by massive central black holes, raised the entropy of the gas already 
collapsed in halos$^{126}$), can 
lead to disk masses, atomic hydrogen distribution  and X-ray emission in better agreement with observations compared to the case
in which the gas density follows the cuspy dark matter profile. Comparing the prediction of simulations to
observations in different wavelengths, such as radio wavelengths at which the atomic hydrogen is revealed, and X-rays by means
of which the hot gas in the halo can be detected, could be a powerful tool to constrain the physics of 
galaxy formation. 

\subsection{Additional issues at a glance}

There are some aspects of the galaxy formation process that we have not covered in this report and that
might have an impact on the problem of disk formation and, in particular,
on the origin of bulgeless galaxies.
First, gas accretion seems to occur in a different fashion for large and
small galaxies. Small galaxies mostly accrete gas that enters already cold ($T \sim 10^4$ K) in the dark matter
halo following a filamentary, anisotropic flow, while large galaxies tend to be built mostly
by accretion of gas that is shock heated to high temperatures ($T \sim 10^6$ K) when it enters the halo and later cools 
down in a smooth, isotropic flow$^{127, 128}$. 
The implications of these two modes of accretion on disk formation, for example 
their impact on angular momentum transport, are still unclear. The treshold halo mass that decides
between cold and hot mode accretion is estimated to be around $10^{12}$ solar masses, namely
comparable to the mass of the Milky Way or Andromeda halo. 
This is interesting in the context of 
bulgeless galaxies. As we already noted, bulge-dominated galaxies such as the Andromeda galaxy, are usually 
much more massive than bulgeless galaxies such as the M33 galaxy (there are some examples of 
massive bulgeless galaxies, e.g. M101 in Figure 1), hence their
different structure might be, at least partially, a product of these two different gas accretion modes.

Another aspect that needs attention is that for galaxies mainly accreting in the hot mode a thermal instability
might develop in the cooling flow$^{129}$.
As a result of that gas cools faster in slightly more overdense regions,
giving rise to dense clouds nearly in pressure equilibrium with the surrounding gas$^{129,130}$ (see Figure 12). 
In this case the hot mode would not be a smooth
cooling flow but would develop into a two-phase medium. Clouds lose orbital angular momentum due
to the hydrodynamical drag exerted by the surrounding gas and eventually reach the center at timescales
different from gas that started out at the same radius but remained in the diffuse phase.
Since there is transfer of angular momentum between the cold and hot phase we expect some
effects on disk formation. It is hard to predict, even qualitatively, in which direction 
the effect will be since both the
hot and the cold phase will eventually contribute to the disk assembly. A very high resolution
is required to resolve the thermal instability$^{130}$ and cosmological simulations are just now
becoming capable to do so.

Furthermore, another issue concerns a dynamical mechanism called adiabatic contraction. 
Indeed the dark matter halo should slowly contract in response to the baryonic mass collapsing
within it$^{131}$. This is a standard assumption of the one dimensional numerical models
that we discussed earlier in this review (see section 1). Different researchers who have developed
numerical models of gas collapse within isolated spherical halos disagree slightly on
the magnitude of the effect, i.e. on how strongly the overall potential well of the
galaxy is modified$^{132}$. In any case, the effect of adiabatic contraction is to increase the maximum rotational
velocity of the galaxy and to reduce the size of the disk
because the overall potential well becomes deeper (hence the radius at which a gas parcel
can be in centrifugal equilibrium diminishes$^{8}$). This has an impact on any observed correlation
between galaxy properties that involves the disk rotational velocity, for example the Tully-Fisher
relation (see previous section). Recent work with one-dimensional numerical models has 
pointed out that, if adiabatic contraction is effective, it is
impossible to match the observed Tully-Fisher relation even in the case that the angular momentum of the gas is
perfectly conserved during the collapse -- the resulting galaxies always rotate too fast$^{13}$.
The fact that recent cosmological simulations are
able to roughly match the Tully Fisher relation$^{55}$ is thus not expected. However,
this discrepancy might suggest that one-dimensional models cannot capture the dynamics and 
thermodynamics of the three-dimensional collapse in a cosmological context.
In particular, it is possible that the concept of adiabatic contraction is not  appropriate
for a structure formation model like $\Lambda$CDM in which a large fraction of the baryonic mass is added
not via slow, smooth accretion of gas but rather via mergers and/or cold flows on timescales short enough
to violate the assumption of adiabaticity in the first place.

Finally, even if realistic disks could form it is not clear that they will survive intact until the present
epoch in a Universe where structure grows hierarchically. In a hierarchical Universe a galaxy
is always surrounded by smaller galaxies that will eventually perturb it during close fly-bies by raising tides, or even merge with it. 
Cosmological simulations predict that these satellite galaxies have very eccentric, plunging  orbits$^{133}$
which should take them close to the disk of the primary galaxy. 
The tidal perturbations will deposit kinetic energy in the stellar disk, raising the random velocities
of its stars and increasing its scale height$^{134}$. This is confirmed by recent
three-dimensional simulations, although the extent of the damage, especially whether or not most
of the disk survives intact despite the intruders, is still debated$^{135,136}$.
Tidal interactions will
also trigger bar formation and eventually produce a bulge via the buckling instability 
(see section 4.1). The disk might eventually re-growth as new gas is accreted from the 
halo or a gas-rich companion is digested, but this is still unclear at the moment.
The study of disk heating by satellites is a difficult problem that requires a resolution
beyond that currently possible in cosmological simulations.

In summary,  there are several aspects of galaxy formation that still need to be understood
in depth, and their overall impact on disk formation thus still awaits a clear assessment.
This deeper understanding demands a substantial improvement in the resolution of the simulations
and in the sub-grid recipes that describe the ISM, star formation and feedback. Yet, in our opinion, simply
more computing power and better sub-grid methods are not going to solve many of the pressing issues in this
field, such as the origin of bulgeless galaxies. Instead, there  is still room, and need, for
new ideas and new approaches to the open questions. Provided that this happens, computer
simulations will then continue to play a central role in advancing our understanding of
how disk galaxies form.

\vskip 2truecm

{\bf{\Large{Acknowledgements}}}

\bigskip

We thank Volker Springel and Elena d'Onghia for reading an earlier version of the manuscript and
returning very helpful comments. We also thank all the people that have had a long term collaboration
with the authors and contributed to many of the results discussed in this review; Alyson Brooks, Chris
Brook, James Wadsley, Joachim Stadel, Tom Quinn, Greg Stinson, Beth Willman, Neal Katz, George Lake and Ben Moore. We 
also acknowledge useful and stimulating discussions on galaxy formation, star formation and computer modeling issues with
James Bullock, Stefano Borgani, Marcella Carollo, Stephane Courteau, Victor Debattista, Avishai Dekel, Aaron Dutton, Andrea Ferrara, 
Nick Gnedin, Dusan Keres, Ralf Klessen, Andrey Kravtsov, Stelios Kazantzidis, Simon Lilly, Aryeh Maller, Francesco Miniati, Julio Navarro, 
Padelis Papadopoulos, Cristiano Porciani, Matthias Steinmetz, Romain Teyssier, Frank van den Bosch and Keichi Wada.

\end{document}